\documentclass{SciPost}

\hypersetup{
    colorlinks,
    linkcolor={red!50!black},
    citecolor={blue!50!black},
    urlcolor={blue!80!black}
}

\usepackage{amsmath,amssymb}
\usepackage{tikz,pgfplots}
\usepgfplotslibrary{fillbetween}
\usetikzlibrary{arrows.meta}
\usepackage{braket}
\usepackage{color}
\definecolor{aone}{rgb}{0.1,0.4,0.2}
\definecolor{atwo}{rgb}{0.1,0.8,0.4}
\definecolor{bone}{rgb}{0,0,0.45}
\definecolor{btwo}{rgb}{0.1,0.1,0.9}
\definecolor{ctwo}{rgb}{1,0.1,0.1}
\definecolor{cone}{rgb}{0.9,0.6,0}
\usepackage{empheq}
\usepackage[most]{tcolorbox}
\newtcbox{\highlightbox}[1][]{nobeforeafter,math upper,tcbox raise base,enhanced,colback=black!5,colframe=black,drop fuzzy shadow,left=0.5em,top=0.4em,right=0.6em,bottom=0.4em}
\usepackage[bitstream-charter]{mathdesign}
\DeclareSymbolFont{usualmathcal}{OMS}{cmsy}{m}{n}
\DeclareSymbolFontAlphabet{\mathcal}{usualmathcal}

\fancypagestyle{SPstyle}{
\fancyhf{}
\lhead{\colorbox{scipostblue}{\bf \color{white} ~SciPost Physics }}
\rhead{{\bf \color{scipostdeepblue} ~Submission }}

\fancyfoot[C]{\textbf{\thepage}}
}

\newcommand{\ci}{\mathrm{i}\mkern1mu}

\newcommand{\bx}{\mathbf{x}}
\newcommand{\by}{\mathbf{y}}
\begin{document}

\pagestyle{SPstyle}

\begin{center}{\Large \textbf{\color{scipostdeepblue}{
Two dimensional inhomogeneous classical systems at criticality\\
}}}\end{center}

\begin{center}\textbf{
Jean-Marie St\'ephan\textsuperscript{1
}
}\end{center}

\begin{center}
{\bf 1} CNRS, ENS de Lyon, Laboratoire de Physique, UMR 5672, F69342 Lyon, France
\\[\baselineskip]
\end{center}

\section*{\color{scipostdeepblue}{Abstract}}
\textbf{\boldmath{%
We study simple inhomogeneous deformations of known two-dimensional classical lattice models described by conformal field theory (CFT) at large distances. The deformations are chosen to vary slowly at lattice scales, while preserving critical behavior. Globally, we find that such systems are described by a CFT in curved space, and identify the underlying space metric. Our two examples are the Ising model and the six vertex model with domain wall boundary conditions. In the latter boundaries are also inhomogeneous, which complicates the analysis. We nevertheless solve the free case using hydrodynamics. In the presence of interactions we determine the arctic curve which separates a critical fluctuating region from an ordered (frozen) phase.
}}

\vspace{10pt}
\noindent\rule{\textwidth}{1pt}
\tableofcontents
\noindent\rule{\textwidth}{1pt}
\vspace{10pt}

\section{Introduction}
\label{sec:intro}
The in-depth study of clean simplified lattice models of statistical physics has a venerable history, in particular in relation to critical phenomena. Owing to their simplicity, these models can at worst be studied numerically efficiently, and a lot can be understood by quantum field theory approaches, where scale invariance typically is promoted to conformal invariance \cite{Polyakov}. 

The situation is particularly favorable in two space dimensions. On the lattice side several relevant spin or vertex models are exactly solvable or integrable\cite{Baxter1982,Gaudin_2014}. On the field theory side the corresponding $2d$ conformal field theories (CFT) \cite{BPZ,Gawedzki_cft,Yellowbook,Mussardo} are well understood. This often allows for the exact determination of critical exponents, and precise understanding of the  correlation functions which are constrained by conformal symmetry.

When comparing to field theories, lattice models are usually chosen to depend on very few parameters: the system is homogeneous, and this leads to a homogeneous CFT, with translation invariance broken only at the boundary or perhaps by localized impurities. Typically homogeneous boundaries renormalize to conformal invariant boundaries \cite{Cardy_1989}, so the CFT is still homogeneous in the bulk. There are examples of boundary conditions that break conformal invariance, for example with domain-wall or emptiness type strongly inhomogeneous boundary conditions in the presence of a $U(1)$ symmetry \cite{Abanov_hydro,Stephan_efp}. In this case the correct description is that of a field theory in curved space, with a nontrivial metric determined from the boundary conditions \cite{ADSV2016}.

In this paper we consider inhomogeneous deformations of critical integrable lattice models without disorder, where the Boltzmann weights are chosen to vary slowly in space. The system is therefore locally homogeneous, but not globally. We perform such a deformation without breaking criticality, and our main findings are twofold: the first is that the metric  becomes space dependent, leading to a field theory description in curved space. Secondly, we will see that such deformations can be made compatible with integrability so are very natural from this perspective as well. Our main examples will be the Ising model and the six vertex model with domain wall boundary conditions. The former best illustrates illustrates the first point, and the latter the second point, while providing an example where both boundary conditions and integrable weights are inhomogeneous.

The paper is organized as follows. In section \ref{sec:Ising} we consider a two-dimensional version of the Ising model and discuss its behavior in the context of free fermions and the inhomogeneous Ising CFT. We then turn our attention to the integrable inhomogeneous six vertex model with domain wall boundary conditions. For technical reasons we divide its investigation into two parts: we consider in section \ref{sec:6vdwfree} the free fermionic case, while the study of the  interacting model, where less can be said, is postponed to section \ref{sec:6vdw}. Before getting to the bulk of the manuscript, we discuss at the end of this introduction a class of related one-dimensional inhomogeneous quantum systems, which serve as additional motivation.

\paragraph{Inhomogeneous quantum systems.}

Consider a simple fermionic tight binding model on the $1d$ line, which can be written as
\begin{align}
	H=-J \sum_{j} \,h_{j,j+1}\,,
\end{align}
where $h_{j,j+1}=(c_{j+1}^\dag c_j+c_j^\dag c_{j+1})/2$ and the $c_j,c_j^\dag$ are fermionic operators satisfying the usual anticommutation relations. In momentum space the fermions have dispersion $\varepsilon(k)=-J\cos k$. With $k_F$ the Fermi momentum ($k_F=\pi/2$ here), the Fermi speed is $v_F=\varepsilon'(k_F)=J$. At low energies, this model is described by a free Dirac fermion theory. In particular light cones propagation is given by $x=\pm v_F t$. Said differently, the underlying space-time metric can be written as $ds^2=dx^2-v_F^2 dt^2$, and the light cones are the corresponding (null) geodesics. 

In many experimentally relevant setups translation invariance is broken by a confining potential $V(x)$, assumed to vary slowly at the lattice scale, see for example \cite{MinguzziGangardt2005,QuinnHaque2014,Dean_2019}. Using the local density approximation, the local Fermi momentum $k_F(x)$ is set by the confining potential, $\varepsilon(k_F(x))+V(x)=0$, and the local Fermi speed $v_F(x)$ is in turn set by the local Fermi momentum. The corresponding fermionic density is $\rho(x)=k_F(x)/\pi$. At low energies, the system is described by the same Dirac theory, albeit in curved space-time, with an underlying metric which can be written as  $ds^2=dx^2-v_F(x)^2dt^2$, see \cite{DSC2017}. 

Another slightly different setting is that of slowly modulated Hamiltonian densities
\begin{align}\label{eq:deformedH}
	H=-J\sum_{j=1}^N f(\frac{j+1/2}{N}) \,h_{j,j+1}.
\end{align}
For such deformations the Fermi momentum $k_F$ stays constant, so density stays constant, however the Fermi speed is modified to $v_F(x)=Jf(x)$. The underlying space-time metric is $ds^2=dx^2-v_F(x)^2dt^2$, and as a consequence light cones are also  curved \cite{curvedlightcones}. This effect was recently simulated on  programmable quantum processors\cite{Rhyno_2026}. Another consequence is a global rescaling of the low energy spectrum above the ground state\cite{Wen_2016}. 

This type of deformed quantum chains has been investigated in several other different contexts, and for particular choices of deformations $f(x)$. For example, the specific choice $f(x)=\sqrt{x(1-x)}$ is known to yield perfect transfer \cite{Christandl_2004_PRL,Albanese_2004,Christandl_2005,Kay_2010} of states in a finite time. This case and other generalizations where single particle energies can be obtained in closed form have been investigated from the perspective of integrability and orthogonal polynomials \cite{Crampe_2019,Bernardetal2025,SSH_2026}. There is also a relation to entanglement and entanglement Hamiltonians of homogeneous systems. Writing $\rho_A=e^{-H_A}$ the reduced density matrix for a subsystem $A$, the entanglement Hamiltonian $H_A$ of critical $1d$ systems in their ground state is of the form \eqref{eq:deformedH} for the choice $f(x)=x(1-x)$, see \cite{BisognanoWichmann,Unruh,Peschel_2009,CardyTonni_2016,Wong2013,KlichVamanWong_2017,Eisler_2019,Eisler_2020}. Finally, the case of sine square deformation has been investigated in depth \cite{Gendiar_2009,Gendiar_2010,Maruyama_2011,Katsura_2011,Katsura_2012}, in part because the ground state with open boundary conditions coincides exactly with the ground state of a periodic homogeneous chain. 

We note that similar deformations can be performed with other models that map to free fermions as well, in particular for the Ising chain in transverse field. The two-dimensional inhomogeneous Ising model discussed in the next section is related to this one, with euclidean space replacing Minkowski space-time. The inhomogeneous six vertex model discussed in sections \ref{sec:6vdwfree} and \ref{sec:6vdw} bears similarities to a deformed Hamiltonian \eqref{eq:deformedH} in a trapping potential, the main difference being that determining the corresponding $k_F(x),v_F(x)$ becomes a highly nontrivial task. 
\section{The two-dimensional inhomogeneous Ising model}
\label{sec:Ising}
The Ising model is arguably the most basic model of statistical mechanics \cite{McCoyWu_book}. It describes in particular ferromagnetism and the behavior of water. In two and three dimensions, it has a second order phase transition at a certain critical temperature, and the corresponding critical point has been much studied in relation to CFT.    

Famously, the 2d case on the square lattice was solved by Onsager \cite{Onsager_1944}, and this exact solution serves as an ideal playground to implement the ideas discussed in the introduction. Our main focus will be on the critical point. We start by recalling some basic facts regarding the homogeneous model, before proceeding with our inhomogeneous deformation. 
\subsection{Reminder on the 2d homogeneous Ising model}
We consider classical spins $\sigma(i,j)$ on a square grid, which can only take values $\pm 1$ at each lattice site. The interaction energy is given by
\begin{align}
	E(\sigma)=-\sum_{i=1}^N \sum_{j=1}^{M-1} \left[J \sigma(i,j)\sigma(i,j+1)+J' \sigma(i,j)\sigma(i+1,j)\right],
\end{align}
where $J,J'$ are the vertical and horizontal coupling constants. There is no magnetic field. The model is considered in the canonical ensemble with partition function $Z=\sum_{\sigma} e^{-\beta E(\sigma)}$ where $\beta=\frac{1}{k_B T}$, $T$ being temperature and $k_B$ Boltzmann's constant. Following tradition we write the dimensionless coupling constants as $K=\beta J$ and $K'=\beta J'$, see figure \ref{fig:2dIsing} below. 
\begin{figure}[htbp]
	\centering\begin{tikzpicture}[scale=0.9]
		\draw[<->,thick] (-1.7,0) -- (-1.7,5);
		\draw (-2.2,2.5) node {$M$};
		\draw[<->,thick] (0,5.7) -- (5,5.7);
		\draw (2.5,6.2) node  {$N$};
		\foreach \x in {0,1,2,3,4,5}{
			\draw[thick] (\x,0) -- (\x,5);
			\draw[thick] (0,\x) -- (5,\x);
			\draw (0.5,-0.5) node {$K'$};
			\draw (-0.5,0.5) node {$K$};
			\draw[densely dashed] (-0.5,\x) -- (0,\x);
			\draw[densely dashed] (5,\x) -- (5.5,\x);
		}
		\fill[blue!75!green!25,opacity=0.4] (0,0) -- (5,0) -- (5,5) -- (0,5) --cycle;
		\draw[->,thick] (6.5,0) arc (-45:45:0.7);
		\draw[->,thick] (6.5,1) arc (-45:45:0.7);
		\draw[->,thick] (6.5,2) arc (-45:45:0.7);
		\draw[->,thick] (6.5,3) arc (-45:45:0.7);
		\draw[->,thick] (6.5,4) arc (-45:45:0.7);
		\draw (7.1,0.5) node {$\mathcal{T}$};
		\foreach \x in {0,1,2,3,4,5}{
			\foreach \y in {0,1,2,3,4,5}{
				\fill (\x,\y) circle (0.1cm);
			}
		}
	\end{tikzpicture}
	\caption{Ising model on an $M\times N$ square grid, with periodic boundary conditions in the horizontal direction for simplicity. All vertical coupling constants (times inverse temperature) are set to $K$, while all horizontal ones are set to $K'$. If $K=K'$ the model is isotropic, otherwise it is not. The transfer matrix $\mathcal{T}$ discussed in section \ref{sec:tm} acts in the vertical direction, and is pictured on the right of the figure.}
	\label{fig:2dIsing}
\end{figure}
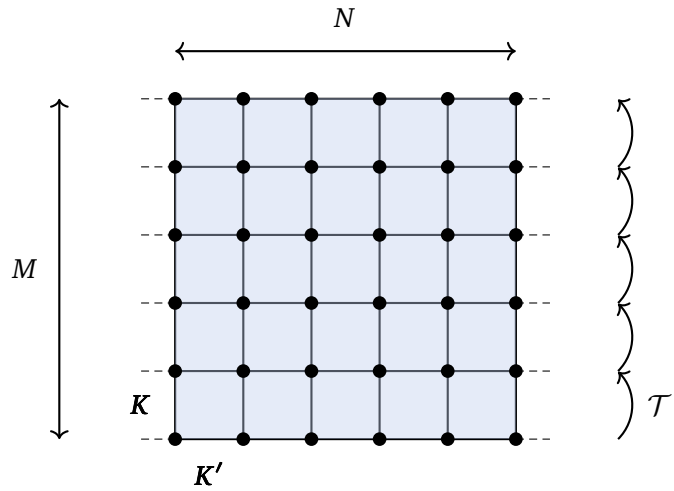

\begin{figure}[htbp]
	\centering
\begin{tikzpicture}
\begin{axis}[xmin=0.05,xmax=1.2,ymin=0.1,ymax=1.2,axis lines=middle,xlabel={$K$},ylabel={$K'$},every axis y label/.style={at={(ticklabel* cs:1.02)},anchor=south},every axis x label/.style={at={(ticklabel* cs:1.02)},anchor=west},axis line style={-{Stealth[scale=2]}},ticks=none]
	\addplot[name path=A,blue,line width=3pt,samples=80,domain=0.05:1.2] {0.5*ln(0.5/sinh(x)+sqrt(1+0.25/(sinh(x)*sinh(x))))};
	\addplot[name path=B,domain=0.05:1.2] {0.1};
	\addplot[name path=C,domain=0.1:1.2] {1.2};
	\addplot[name path=Bbis,domain=0.05:0.1] {0.1};
	\addplot[name path=Cbis,domain=0.05:0.1] {1.3};
	\addplot[green,opacity=0.2] fill between [of=B and A,soft clip={domain=0.05:1.2}];
	\addplot[red,opacity=0.2] fill between [of=A and C,soft clip={domain=0.1:1.2}];
	\addplot[only marks,mark=*,mark size=4pt,point meta=explicit symbolic] coordinates {(0.4407,0.475) [B]};
\end{axis}
\draw (3.1,2.3) node {isotropic};
\draw (3.1,1.95) node {point};
\draw[color=red!50!black] (4.5,4.5) node {\Large{Ordered phase}};
\draw[color=green!50!black] (1.5,0.9) node {\Large{Disordered}};
\draw[color=green!50!black] (1.5,0.4) node {\Large{phase}};
\end{tikzpicture}
\caption{Phase diagram of the two-dimensional Ising model. The critical line \eqref{eq:criticalitycondition} shown in blue separates an ordered phase from a disordered phase. Recall temperature is implicit in this diagram, since $K=\beta J$ and $K'=\beta J'$. All Ising models with parameters on the blue critical line renormalize to the same Ising conformal field theory, so it is not a line of critical points in the field-theoretical sense.}
\label{fig:phasediagram}
\end{figure}
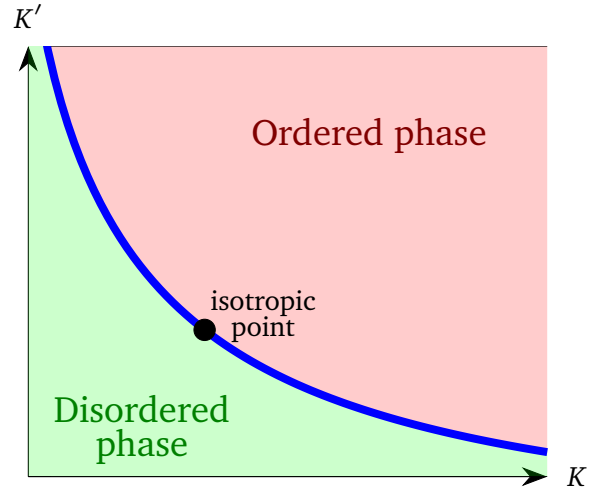

For simplicity we assume in the following that both $M$ and $N$ are even, and periodic boundary conditions (PBC) are imposed in the horizontal direction. 
As was shown by Kramers and Wannier\cite{KramersWannier}, there is a duality which relates the low temperature expansion to the high temperature expansion, allowing the determination of the critical point. It is given by the relation
\begin{align}\label{eq:criticalitycondition}
	K^*=K',
\end{align}
where $K^*$ is defined from $K$ through
\begin{align}\label{eq:star}
	\sinh 2K^*\sinh 2K=1.
\end{align}
Away from this the model is either in a low temperature symmetry broken ordered phase, or in a high temperature disordered phase. If one sticks to critical behavior, we have a one parameter family of critical Ising models indexed (say) by $K'$, with $K$ set by the condition \eqref{eq:criticalitycondition}, see figure \ref{fig:phasediagram} for an illustration. In the scaling limit all of them are described by the \emph{same} Ising conformal field theory, however the limiting procedure going from the lattice to the field theory depends on $K'$. This observation will play an important role in the following. 
\subsection{Transfer matrix and conformal field theory}
\label{sec:tm}
We recall here the facts we need regarding Onsager's transfer matrix solution, and the results which can be deduced from it. We roughly follow the review paper \cite{SchultzMattisLieb}, and refer to appendix \ref{app:IsingTMfermions} and the aforementioned paper for more thorough explanations. 

The transfer matrix is given by
\begin{align}
\mathcal{T}&=\prod_{j=1}^N \left(e^K +e^{-K}\sigma_j^x\right)\;\; \exp\left(K'\sum_{j=1}^N \sigma_j^z \sigma_{j+1}^z\right)\\\label{eq:tmbis}
&=C \exp\left(K^*\sum_{j=1}^N \sigma_j^x\right)\exp\left(K'\sum_{j=1}^N \sigma_j^z \sigma_{j+1}^z\right),
\end{align}
with the usual (quantum) Pauli matrices, the classical spins being identified for now with the two eigenvalues $\pm 1$ of the operator  $\sigma_j^z$.  The two exponentials in \eqref{eq:tmbis} do not commute. $C=[2\sinh 2K]^{N/2}$ is a constant.

The transfer matrix can be used to efficiently compute physical observables, the simplest of which, $Z$, reads for example $Z=\textrm{Tr}\, \mathcal{T}^M$ if PBC are also imposed in the vertical direction. Performing a Jordan-Wigner transformation \eqref{eq:JordanWigner1}, \eqref{eq:JordanWigner2} combined with \eqref{eq:productofexp} brings the transfer matrix to the fermionic form
\begin{align}
	\mathcal{T}=C\exp\left(\frac{1}{2} t_{ij}f_i^\dag f_j\right),
\end{align}
where $f_i^\dag=c_i^\dag$, $f_{i+N}^\dag=c_i$ for $i=1,2,\ldots,N$ and Einstein summation over repeated indices is implied. As usual fermions obey the anticommutation relations $\{c_i,c_j^\dag\}=\delta_{ij}$ and $\{c_i,c_j\}=0$. $t=t_{ij}$ is a $2N\times 2N$ matrix discussed in the  appendix. Hence the transfer matrix is essentially the exponential of a quadratic fermion Hamiltonian. The quadratic Hamiltonian can be diagonalized by standard means, and the spectrum determined accordingly. Care must be taken with boundary conditions. For example the periodic boundary conditions for spins translate into two possible sectors for the fermions, we refer again to appendix \ref{app:IsingTMfermions} for a thorough discussion of all these aspects.

\paragraph{Gap of the transfer matrix.} Denote by $\Lambda_0$ the largest eigenvalue of the transfer matrix, and by $\Lambda_1$ the second largest one in the same sector. The "gap" $\Delta E=\log \frac{\Lambda_0}{\Lambda_1}$ at criticality scales for large $N$ as
\begin{align}
	\Delta E=\frac{2\pi v_F c}{N}+\ldots
\end{align}
where 
\begin{align}
\label{eq:fermispeedIsing}
v_F=\sinh 2K'
\end{align} plays the role of a Fermi speed, and $c=1/2$. This result coincides with the predictions of CFT \cite{Yellowbook,Mussardo} with $c$ the universal central charge. The $\dots$ denote subleading corrections as $N\to\infty$. If one interprets the vertical direction as imaginary time, the underlying space-time metric can be written as $ds^2=dx^2+v_F^2 d y^2$. Note that for the isotropic model $v_F=1$, consistent with the fact that there should be no distinction between the two space coordinates both on the lattice and in the scaling limit.

\paragraph{Spin correlation functions.}
Another important result deals with the behavior of (equal imaginary time) spin-spin correlations, which can, using free fermions techniques, be expressed as a determinant. In the infinite cylinder geometry ($M\to\infty$) this determinant has been much studied in the mathematical literature (e.g. \cite{Isingimpetus}) following the unpublished work of Kaufman-Onsager \cite{Baxter_Onsagerhistory}. Combining available results \cite{KaufmanOnsager,Yang1952,Fisher_1954,Wu_1966,Wuetal_1976} with basic consequences of conformal invariance yields
\begin{align}\label{eq:spinspincorelations}
	\braket{\sigma(m,y)\sigma(m+r,y)}\sim \frac{2^{1/12}e^{3\zeta'(-1)}[\cosh 2K']^{1/4}}{\left[\frac{N}{\pi}\sin \frac{\pi r}{N}\right]^{1/4}}
\end{align}
for $r,N-r,N \gg 1$. The numerical prefactor $2^{1/12}e^{3\zeta'(-1)}\simeq 0.64500$ is non universal, and does not play a significant role in the following. One also recognizes the celebrated Kaufman-Onsager-Yang magnetization exponent $1/4$ in this expression.  
\subsection{Inhomogeneous Ising model}
\begin{figure}[htbp]
	\centering
		\begin{tikzpicture}[scale=1]
			\foreach \x in {0,1,2,3,4,5}{
				\draw[thick] (\x,0) -- (\x,5);
				\draw[thick] (0,\x) -- (5,\x);
				\draw (-0.5,-0.45) node {$K'_{1/2}$};
				\draw (0.5,-0.45) node {$K'_{3/2}$};
				\draw (1.5,-0.45) node {$K'_{5/2}$};
				\draw (2.5,-0.45) node {$K'_{7/2}$};
				\draw (3.5,-0.45) node {$K'_{9/2}$};
				\draw (4.65,-0.45) node {$K'_{11/2}$};
				\draw (-0.25,0.5) node {$K_1$};
				\draw (0.75,0.5) node {$K_2$};
				\draw (1.75,0.5) node {$K_3$};
				\draw (2.75,0.5) node {$K_4$};
				\draw (3.75,0.5) node {$K_5$};
				\draw (4.75,0.5) node {$K_6$};
				\draw[densely dashed] (-0.5,\x) -- (0,\x);
				\draw[densely dashed] (5,\x) -- (5.5,\x);
			}
			\fill[blue!90!white,opacity=0.4] (0,0) -- (0.5,0) -- (0.5,5) -- (0,5) --cycle;
			\fill[blue!75!white,opacity=0.4] (0.5,0) -- (1,0) -- (1,5) -- (0.5,5) --cycle;
			\fill[blue!60!white,opacity=0.4] (1,0) -- (1.5,0) -- (1.5,5) -- (1,5) --cycle;
			\fill[blue!45!white,opacity=0.4] (1.5,0) -- (2,0) -- (2,5) -- (1.5,5) --cycle;
			\fill[red!30!blue,opacity=0.3] (2,0) -- (2.5,0) -- (2.5,5) -- (2,5) --cycle;
			\fill[red!50!blue,opacity=0.4] (2.5,0) -- (3,0) -- (3,5) -- (2.5,5) --cycle;
			\fill[red!60!blue,opacity=0.3] (3,0) -- (3.5,0) -- (3.5,5) -- (3,5) --cycle;
			\fill[blue!60!red,opacity=0.3] (3.5,0) -- (4,0) -- (4,5) -- (3.5,5) --cycle;
			\fill[blue!70!red,opacity=0.4] (4,0) -- (4.5,0) -- (4.5,5) -- (4,5) --cycle;
			\fill[blue!85!red,opacity=0.4] (4.5,0) -- (5,0) -- (5,5) -- (4.5,5) --cycle;
			\draw[->,thick] (6.5,0) arc (-45:45:0.7);
			\draw[->,thick] (6.5,1) arc (-45:45:0.7);
			\draw[->,thick] (6.5,2) arc (-45:45:0.7);
			\draw[->,thick] (6.5,3) arc (-45:45:0.7);
			\draw[->,thick] (6.5,4) arc (-45:45:0.7);
			\draw (7.3,0.5) node {$T$};
			\foreach \x in {0,1,2,3,4,5}{
				\foreach \y in {0,1,2,3,4,5}{
					\fill (\x,\y) circle (0.08cm);
				}
			}
	\end{tikzpicture}
	\caption{2d inhomogeneous Ising model. The coupling constant are space dependent in the horizontal direction, but still homogeneous in the time direction, as illustrated by the color code. As a consequence, there is still only one transfer matrix.}
	\label{fig:2dIsinginh}
\end{figure}
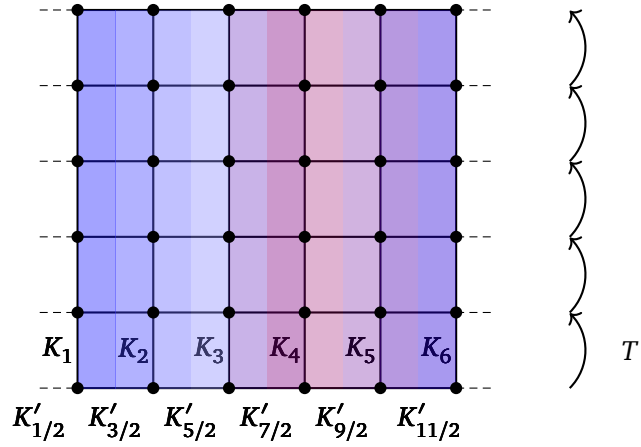

As is well-known (see e.g. \cite{Au-YangMcCoy_1974p1,Au-YangMcCoy_1974p2,PhysRevE.88.032147,McCoyWu_book,Chelkak_2024} where this or similar deformations sometimes go under the name layered Ising model), the transfer matrix can still be constructed  for space-dependent coupling constants $K\to K_j$, $K'\to K'_{j+1/2}$. The so-generalized transfer matrix reads
\begin{align}\label{eq:generalIsingTM}
	\mathcal{T}=C\exp\left(\sum_{j=1}^N K_j^* \sigma_j^x\right)\exp\left(\sum_{j=1}^N K'_{j+1/2}\sigma_j^z \sigma_{j+1}^z\right),
\end{align}
where now $C=\prod_{j=1}^N \sqrt{2\sinh 2K_j}$. There is still a mapping to a (more complicated) free fermion model through a Jordan-Wigner transformation \eqref{eq:JordanWigner1}, \eqref{eq:JordanWigner2} combined with equation \eqref{eq:productofexp}, see appendix \ref{app:IsingTMfermions} for details. The underlying model is illustrated in figure \ref{fig:2dIsinginh}.

We are looking for the simplest inhomogeneous modification of the model which leads to a CFT with a non trivial space metric. To find it, one needs a deformation which varies slowly on lattice scales. This guarantees the existence of a mesoscopic scale for which the system looks homogeneous, while still being inhomogeneous as a whole. The arguably most natural way is to make the inhomogeneous coupling constants space-dependent as follows
\begin{align}\label{eq:choice1}
	(K_j)^* &=K_0\, f\left(\frac{j}{N}\right),\\ \label{eq:choice2}
	K'_{j+1/2}&=K_0\, f\left(\frac{j+1/2}{N}\right),
\end{align}
where $(K_j)^*$ is defined from $K_j$ through \eqref{eq:star}. $K_0$ is the isotropic coupling constant $K_0=\frac{\log(1+\sqrt{2})}{2}$ and $f$ is a sufficiently reasonable function which does not vanish. If we assume periodic boundary conditions in the horizontal direction then $f$ should be periodic with period one. The choice \eqref{eq:choice1}, \eqref{eq:choice2} maintains local critical behavior since $K_j^*\simeq K'_{j+1/2}$ for large $N$. Said differently, by separation of scales the inhomogeneous model globally explores the critical parameter line shown in figure \ref{fig:phasediagram}, while behaving locally like a given point on this line \eqref{eq:criticalitycondition}.
\subsection{Curved space-time description of the inhomogeneous model}
For our choice of inhomogeneous deformation \eqref{eq:choice1}, \eqref{eq:choice2}, the exact formula \eqref{eq:fermispeedIsing} in the homogeneous model strongly suggests a local Fermi speed of the form
\begin{align}
	v_F(x)=\sinh [2K_0 f(x/N)].
\end{align}
In field theory language the underlying euclidean space-time metric can be written as
\begin{align}
	ds^2=dx^2 + v_F(x)^2 dy^2.
\end{align}
Introducing the curved coordinates 
\begin{align}\label{eq:newcoordinate}
	\widetilde{x}=\int_0^x \frac{du}{v_F(u)}
\end{align}
yields the equivalent form
\begin{align}
	ds^2=v_F(x)^2 \left[d\widetilde{x}^2+dy^2\right].
\end{align}
These so-called isothermal coordinates are conceptually nicer, and simplify most computations. It is also sometimes convenient to use complex coordinates $z=\widetilde{x}+\ci y$, in which case
\begin{align}\label{eq:Isingdzdzbar}
	ds^2=e^{2\sigma} dz d\bar{z}
\end{align} 
with Weyl factor $e^{\sigma}=v_F(x)$. 

We have thus found an elementary way of changing the underlying space-time metric, by slowly changing the coupling constants of the model as in \eqref{eq:choice1}, \eqref{eq:choice2}. Since  we work in euclidean space it is not possible to probe it by investigating light cone propagation as in the quantum setting. Nevertheless, the change of metric has interesting consequences on the energy spectrum and physical observables. We discuss below two of those, namely the gap of the transfer matrix and the behavior of spin-spin correlation functions.

\paragraph{Gap of the transfer matrix.}
Well-known CFT reasoning predicts (see e.g. \cite{Wen_2016} for a similar quantum setting discussed in the introduction) a gap
\begin{align}
	\Delta E= \frac{2\pi c}{\widetilde{N}}+\ldots
\end{align}
where $c$ is the central charge, and $\widetilde{N}=\int_0^N \frac{dx}{\sinh 2K_0 f(x/N)}$ acts as an effective system size. By gap what we really mean is the gap in the ground state sector, so in CFT parlance we only consider the conformal tower corresponding to the identity operator\footnote{The first excited state of the full transfer matrix lies in the other sector. We checked that this true gap is consistent with known CFT data as well.}.
For generic deformations, it is not obvious how to calculate the gap exactly in finite size. Nevertheless, the free fermion mapping still yield values for large $N$ to arbitrary numerical precision, which allows to check the prediction. Some results are collected in table \ref{tab:gap} and show excellent agreement.
\begin{figure}[htbp]
	\centering\begin{tabular}{|c|c|c|c|}
		$N$&$\widetilde{N}\Delta E^{(0)}/2\pi$&$\widetilde{N}\Delta E^{(1)}/2\pi$&$\widetilde{N}\Delta E^{(2)}/2\pi$\\
		\hline
		32& 0.499598&0.489202 & 0.471669\\
		64& 0.499899& 0.498269& 0.494033\\
		128& 0.499974& 0.499573& 0.498586\\
		256& 0.499993& 0.499893& 0.499650\\
		512& 0.499998& 0.499973& 0.499912\\
		\hline
	\end{tabular}
	\caption{Numerical values for the gap $\Delta E^{(p)}$ for the choice of deformation $f(u)=1+a\sin 2\pi p u+b\sin 6\pi p u$. Here $a=0.7$, $b=0.5$. The case $p=0$ corresponds to the homogeneous Ising model. 
	}
	\label{tab:gap}
\end{figure} 
Let us nevertheless mention that the result requires some fine-tuning in the inhomogeneous setup, compared to the homogeneous setting. By that we mean that it is important to choose a deformation where $f$ is evaluated at $\frac{j}{N}$ on vertical links and $\frac{j+1/2}{N}$ on horizontal ones, as was done here. Otherwise the universal CFT gap appears to be polluted by non universal terms which are of the same order. Such issues are less present in the quantum setting, possibly because the local Fermi speed is modified linearly, while in the 2d Ising model it enters nonlinearly through equation \eqref{eq:fermispeedIsing}. We study next spin-spin correlation functions, which are less sensitive to this issue, so should be seen as a better check of our inhomogeneous Ising CFT predictions in a statistical mechanics context. 
\paragraph{Long range spin correlation functions.}
  Denote by $\boldsymbol{\sigma}$ the Ising spin field in the underlying CFT. It has dimension $1/8$. The space-time metric can be brought to a trivial form $ds^2 =e^{2\sigma}[d\widetilde{x}^2+dy^2]\to d\widetilde{x}^2+dy^2=dz d\bar{z}$ through a Weyl rescaling. Using the transformation law of primary fields under such Weyl rescaling (see  \cite{Gawedzki_cft}), the two point function reads
  \begin{align}\label{eq:spinspincorelationsinh}
  	\braket{\boldsymbol{\sigma}(x,y)\boldsymbol{\sigma}(x',y)}=\frac{[v_F(x/N)v_F(x'/N)]^{-1/8}}{\left[\frac{\widetilde{N}}{\pi}\sin \frac{\pi(\widetilde{x}-\widetilde{x}')}{\widetilde{N}}\right]^{1/4}},
  \end{align} 
  once again at equal imaginary time in the infinite cylinder geometry. Notice it is no longer translation invariant, even in terms of the stretched coordinates $\widetilde{x},\widetilde{x}'$. 
  
\begin{figure}[htbp]
  \includegraphics[width=0.5\textwidth]{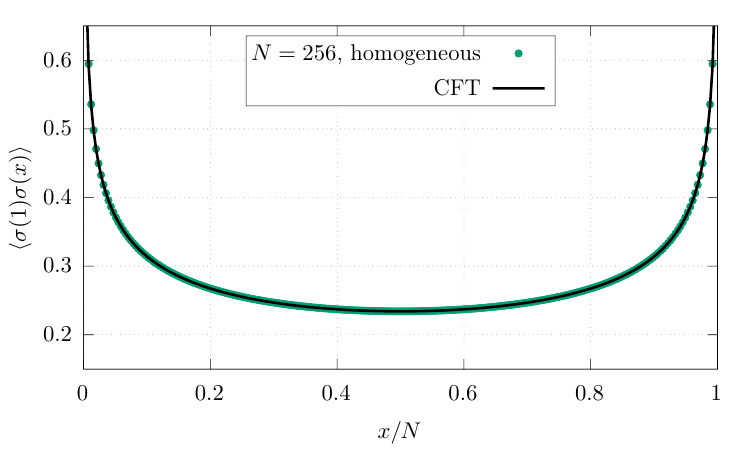}\hfill
  \includegraphics[width=0.5\textwidth]{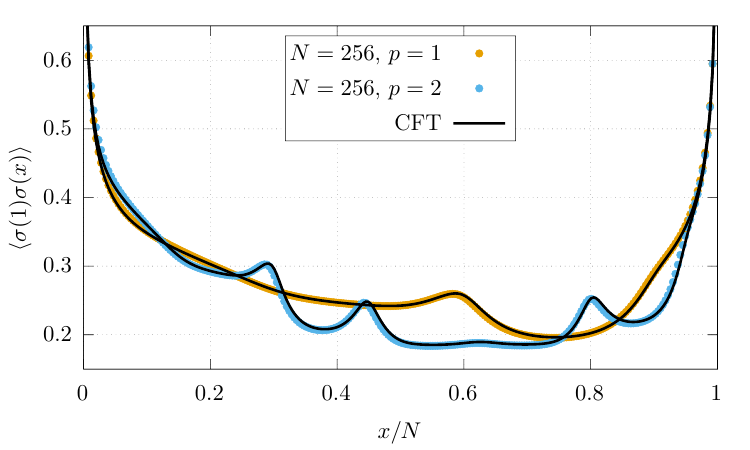}
  	\caption{Equal imaginary time spin-spin correlations for $N=256$ spins, with notation $\sigma(x)=\sigma(x,y=0)$. Left: homogeneous case with known chord length scaling \eqref{eq:spinspincorelations}. Right: inhomogeneous case, with deformation $f(u)=1+a\sin 2\pi p u+b\sin 6\pi p u$,  $a=0.7$, $b=0.5$. Data for $p=1$ and $p=2$ is shown, and it compares extremely well with the curved spacetime CFT result \eqref{eq:isingcurvedspinspin}.}
  	\label{fig:spininh}.
  \end{figure}

  Care must be taken when comparing with the lattice result, as closer inspection of the homogeneous formula \eqref{eq:spinspincorelations} shows a prefactor $[\cosh 2K']^{1/4}$. This suggests the following relation between the lattice spin operator $\sigma(x,y)$ and its field-theoretical counterpart $\boldsymbol{\sigma}$,
  \begin{align}
  	\sigma(x,y)\sim 2^{1/24}e^{(3/2)\zeta'(-1)}(\cosh 2K_0 f(x/N))^{1/8}\boldsymbol{\sigma}(x,y),
  \end{align}
  in the inhomogeneous model (a similar logic has already appeared in  quantum settings, e.g. \cite{RBD2019,BrunDubail}). Therefore the correct prediction is
  \begin{empheq}[box=\highlightbox]{align}
  \label{eq:isingcurvedspinspin}
  	\braket{\sigma(x,0)\sigma(x',0)}\sim 2^{1/12}e^{3\zeta'(-1)} \left[\frac{\cosh [2K_0 f(x/N)]\cosh [2K_0 f(x'/N)]}{v(x/N)v(x'/N)\left(\frac{\widetilde{N}}{\pi}\sin \frac{\pi(\widetilde{x}-\widetilde{x}')}{\widetilde{N}}\right)^2}\right]^{1/8}.
  \end{empheq}
  Equation \eqref{eq:isingcurvedspinspin} is checked numerically in figure \ref{fig:spininh}, with excellent agreement between lattice and CFT calculations.
  
  \paragraph{Discussion.} We have shown how the critical Ising model can simply be deformed to allow for non trivial space-time metric, similar to what occurs in inhomogeneous quantum systems. For simplicity of exposition (rather than technical simplicity) we have considered only the deformation shown in figure \ref{fig:2dIsinginh}, but other choices are possible as well, since for example correlations along the diagonal $x=y$ are known to be somewhat simpler to study exactly, see \cite{Wu_1966,McCoyWu_book,Baxter1982,Chelkak_2024}.
  
  Our findings are not limited to the Ising model. Indeed what one only needs is a one-parameter family of anisotropic models which are still critical and described by the same CFT in the scaling limit, as illustrated in figure \ref{fig:phasediagram}. By making the anisotropy vary slowly in space, criticality is not broken and flows to different theories are prevented. Then, only the underlying space-time metric can change. As an example, our findings can be straightforwardly extended to the (interacting) three-state Potts model, which also has a one-parameter anisotropic critical version which is known exactly. 
  \pagebreak
\section[Free inhomogeneous six vertex model with domain wall boundaries]{Free six vertex model with domain wall boundaries}
\label{sec:6vdwfree}
\begin{figure}[htbp]
\centering	
\begin{tikzpicture}[scale=1.35]
	\foreach \x in {0,2,4,6,8,10}{
		\draw[thick] (-0.5+\x,0) -- (0.5+\x,0);
		\draw[thick] (0+\x,-0.5) -- (0+\x,0.5);
	}
	\draw (0,-1) node {$a_1=a$};\draw (2,-1) node {$a_2=a$};
	\draw (4,-1) node {$b_1=b$};\draw (6,-1) node {$b_2=b$};
	\draw (8,-1) node {$c_1=c$};\draw (10,-1) node {$c_2=c$};
	
\draw[line width=4pt,rounded corners=3mm] (2,-0.5) -- (2,0) -- (2.5,0);
\draw[line width=4pt,rounded corners=3mm] (1.5,0) -- (2,0) -- (2,0.5);
\draw[line width=4pt] (4,-0.5) -- (4,0.5);
\draw[line width=4pt] (5.5,0) -- (6.5,0);
\draw[line width=4pt,rounded corners=3mm] (8,-0.5) -- (8,0) -- (8.5,0);
\draw[line width=4pt,rounded corners=3mm] (9.5,0) -- (10,0) -- (10,0.5);
\begin{scope}[yshift=-2cm]
	\foreach \x in {0,2,4,6,8,10}{
		\draw[thick] (-0.5+\x,0) -- (0.5+\x,0);
		\draw[thick] (0+\x,-0.5) -- (0+\x,0.5);
	}
	\fill[xshift=0.25cm,yshift=0.25cm,color=aone] (-0.5,-0.5) -- (0,-0.5) -- (0,0) -- (-0.5,0) -- cycle;
	\fill[xshift=2.25cm,yshift=0.25cm,color=atwo] (-0.5,-0.5) -- (0,-0.5) -- (0,0) -- (-0.5,0) -- cycle;
	\fill[xshift=4.25cm,yshift=0.25cm,color=bone] (-0.5,-0.5) -- (0,-0.5) -- (0,0) -- (-0.5,0) -- cycle;
	\fill[xshift=6.25cm,yshift=0.25cm,color=btwo] (-0.5,-0.5) -- (0,-0.5) -- (0,0) -- (-0.5,0) -- cycle; 	
	\fill[xshift=8.25cm,yshift=0.25cm,color=cone] (-0.5,-0.5) -- (0,-0.5) -- (0,0) -- (-0.5,0) -- cycle;
	\fill[xshift=10.25cm,yshift=0.25cm,color=ctwo] (-0.5,-0.5) -- (0,-0.5) -- (0,0) -- (-0.5,0) -- cycle;
\end{scope}
\end{tikzpicture}
\caption{Top figure: representation of the six allowed vertices $a_1,a_2,b_1,b_2,c_1,c_2$ of the model. The paths can be seen as particle trajectories which can only go to the right, or upwards. They  can meet at a given vertex but cannot cross. We also consider a case where vertex weights are particle-hole symmetric, as shown.  Bottom figure: color code used in several of the following figures to represent vertex configurations.}	
\label{fig:vertices}
\end{figure}
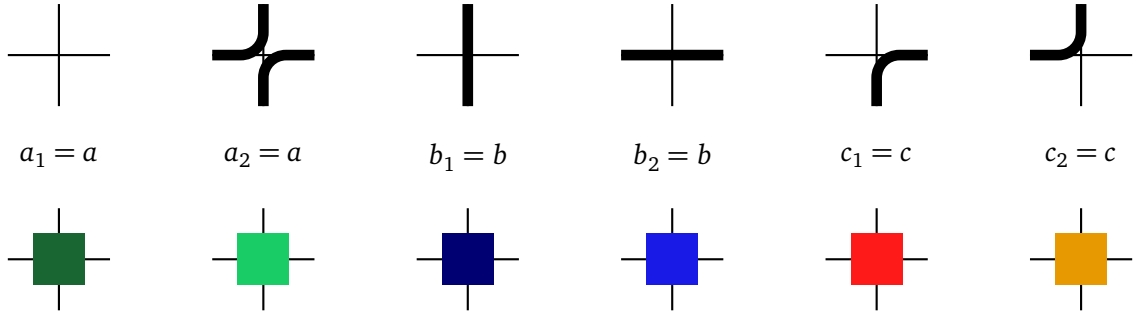

We discuss here a different example, for which the determination of the underlying field theory and density profile requires more work. 
The starting point is the integrable six vertex model, which is a model for lattice paths on the square lattice \cite{Baxter1982,Gaudin_2014}. The paths can only go upwards or to the right, and they can meet but not cross at a given vertex, as is shown in figure \ref{fig:vertices}. In this paper we also restrict to a particle-hole symmetric case with only three independent weights $a,b,c$ remaining. 

It is important to realize that there is a $U(1)$ symmetry in this model, justifying our talking about density later on. A simple way to see it is to rotate the lattice by $45$ degrees, in which case the number of paths (or particles) is conserved going upwards. The total particle number is typically set by boundary conditions. We start by discussing general features of the inhomogeneous and homogeneous models, before specifying the free fermionic case. 
\subsection{Inhomogeneous weights}
From an integrability perspective, a fundamental role is played by an inhomogeneous generalization where the vertex weights depend on horizontal index $j$ and vertical index $k$, in such a way that the anisotropy parameter
\begin{align}
	\Delta=\frac{a_{jk}^2+b_{jk}^2-c_{jk}^2}{2 a_{jk}b_{jk}}
\end{align}
remains constant at every vertex. $\Delta$ parametrizes interactions: the case $\Delta=0$ can be mapped to a system of free fermions, while non constant $\Delta$ breaks lattice integrability. We will consider the following parametrization:
\begin{align}\label{eq:weights}
	a_{jk}=\sin(\lambda_j-\mu_k+\gamma) \qquad,\qquad b_{jk}=\sin(\lambda_j-\mu_k) \qquad,\qquad c_{jk}=c=\sin \gamma
\end{align}
where $\Delta=\cos\gamma$. In this paper, we only study the rational case $|\Delta|<1$, and impose that all weights be positive. Notice $c$ has nothing to do with the central charge discussed before. 
\paragraph{Homogeneous boundary conditions.}
With the results of section \ref{sec:Ising} in mind, it is rather easy to deform the model to get a non trivial space-time metric, while still preserving integrability. The simplest way is to set the rapidities to be of the form $\lambda_j=\lambda(j/N)$, $\mu_k=\mu(k/M)$ say on a $M\times N$ lattice with homogeneous boundary conditions (see also  \cite{Keating2022} for a similar scaling). 

From a field theory perspective the model with homogeneous weights is described by a free compact field with unit central charge (or Luttinger liquid theory) \cite{Yellowbook,Mussardo}. The boson radius (Luttinger parameter) is a function of both particle density and anisotropy $\Delta$. Now with inhomogeneous integrable weights one gets a free field in curved space time with constant Luttinger parameter $K$, which is a CFT, similar to what occurs for slowly modulated Hamiltonian densities, see \cite{curvedlightcones} for a discussion. Hence with homogeneous boundary conditions, integrability preserving weights yield a theory in curved space-time, but with constant $K$. 

\begin{figure}[htbp]
\includegraphics[width=0.33\textwidth]{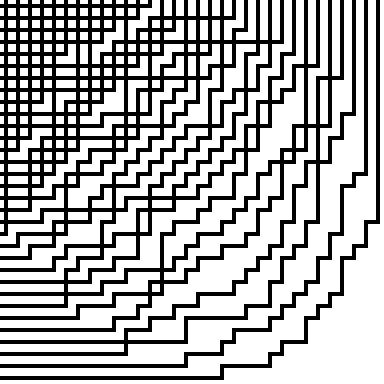}
\includegraphics[width=0.33\textwidth]{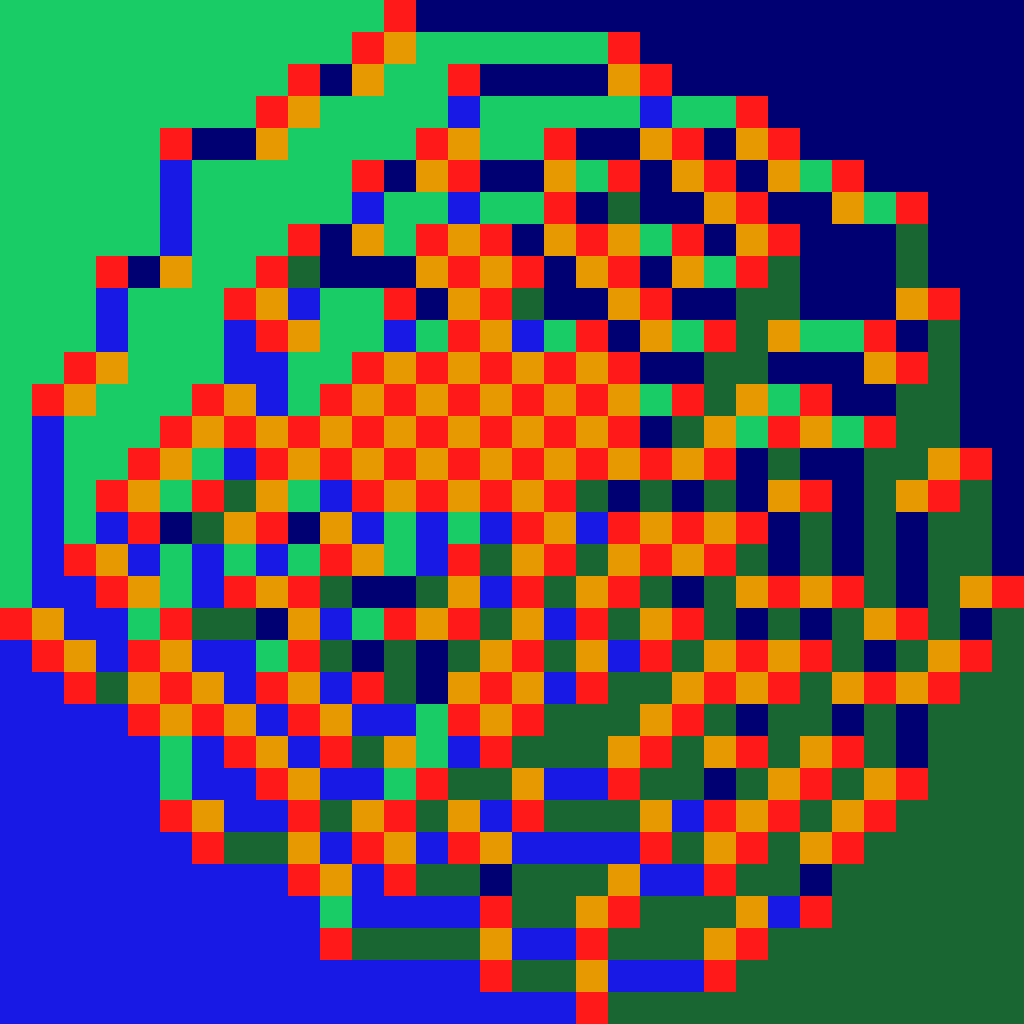}
\includegraphics[width=0.33\textwidth]{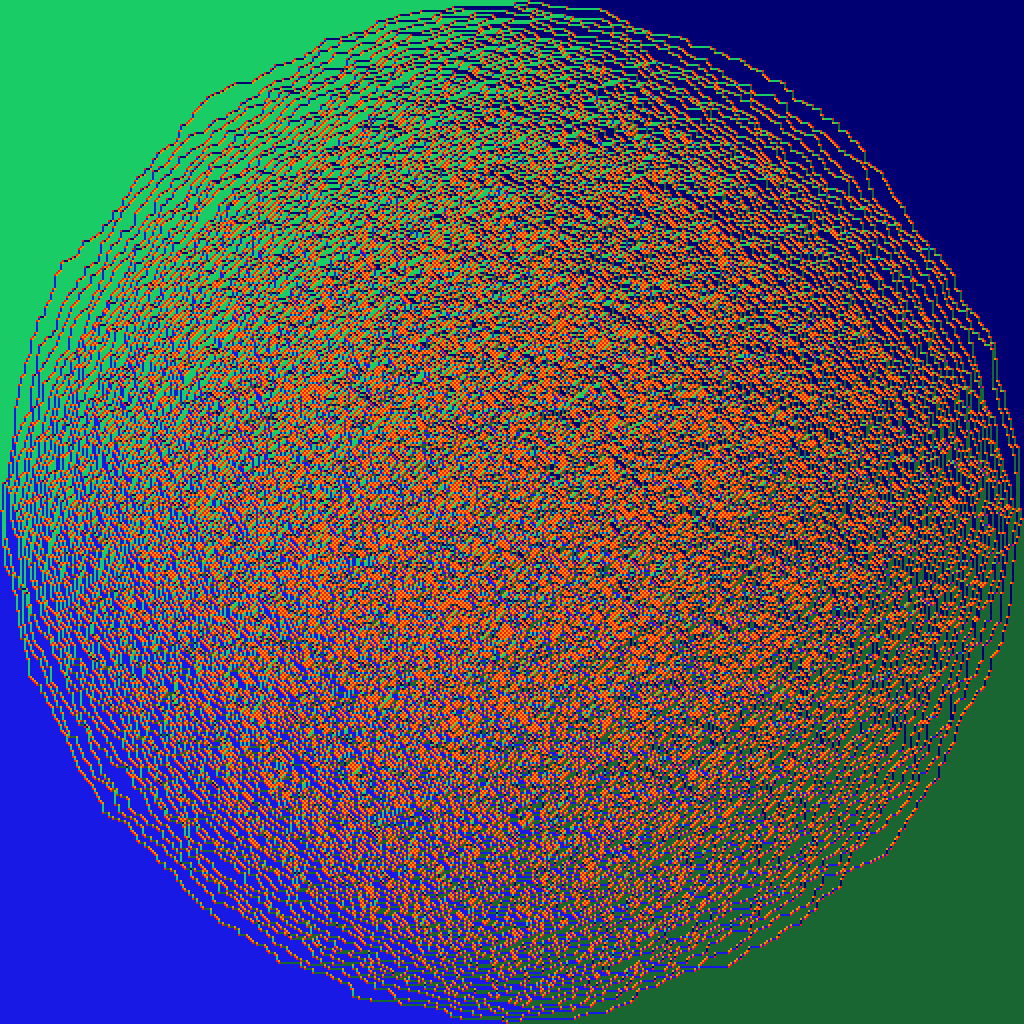}
	\caption{Typical configurations of the six vertex model with domain wall boundary conditions. \emph{Left:} path configurations for $N=32$, which enter through the left and exit through the top of the square. \emph{Middle:} vertex representation for the same configuration, with the color code described in figure \ref{fig:vertices}. \emph{Right:} vertex representation for $N=512$, with "arctic circle" boundary between the fluctuating and frozen regions.}
	\label{fig:6vdwbc}
\end{figure}

Since the situation discussed above is in a sense too similar to that of section \ref{sec:Ising}, we will investigate in the following the richer case of inhomogeneous weights with the simplest strongly inhomogeneous boundary conditions. For general $\Delta$ this study is challenging. For this reason we set $\Delta$ to its free fermionic  value $\Delta=0$ in the remainder of the section, postponing a partial treatment of the interacting case to section \ref{sec:6vdw}.    
\paragraph{Domain wall boundary conditions.} This refers to a particular type of boundary conditions on a $N\times N$ square lattice, with all paths entering from the left boundary, and exiting through the top  boundary. A typical configuration for homogeneous weights is shown in figure \ref{fig:6vdwbc}.

 The domain wall six vertex model has received considerable attention in the homogeneous case \cite{zinnjustin2002,BogoliubovPronkoZvonarev,colomo2010arctic,ADSV2016,TangentMethod}, as it exhibits a limit shape phenomenon (e.g.  \cite{Stephan_lectures}) with spatial phase separation between a frozen (deterministic) phase, and a phase with non trivial density and fluctuations, see figure \ref{fig:6vdwbc}. The boundary between these two phases is called \emph{arctic curve}. There is also an exact mapping to a dimer model on the so-called Aztec diamond, for which it was proven that the arctic curve separating the two phases is a circle \cite{ArcticCircle}. 

Importantly, there is a field theory description for the fluctuating region in terms of a Dirac theory in curved space-time \cite{ADSV2016} or a free field (Luttinger liquid theory) \cite{kenyon2007}. The difference with the results of the previous section is that the non trivial space-time metric is generated by inhomogeneous boundary conditions, rather than inhomogeneous Boltzmann weights.

\begin{figure}[htbp]
\includegraphics[width=0.33\textwidth]{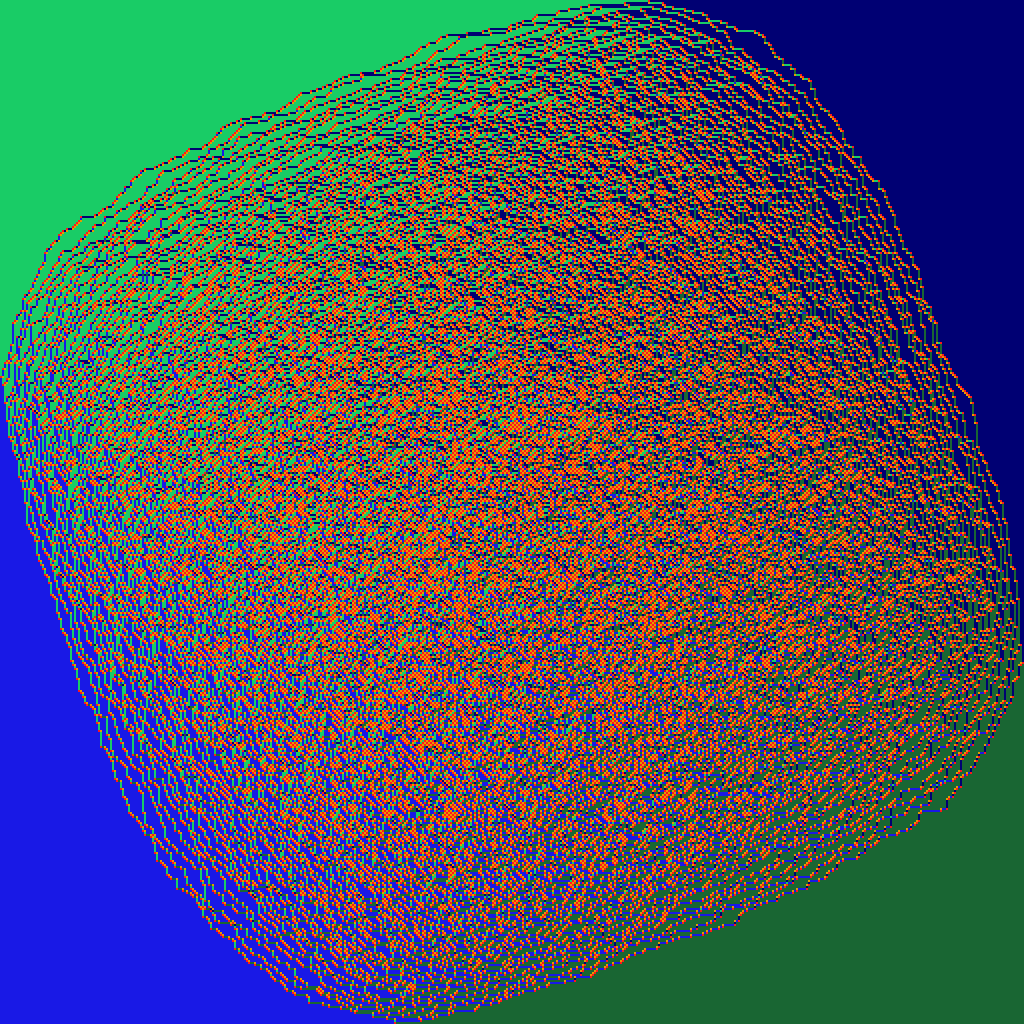}
\includegraphics[width=0.33\textwidth]{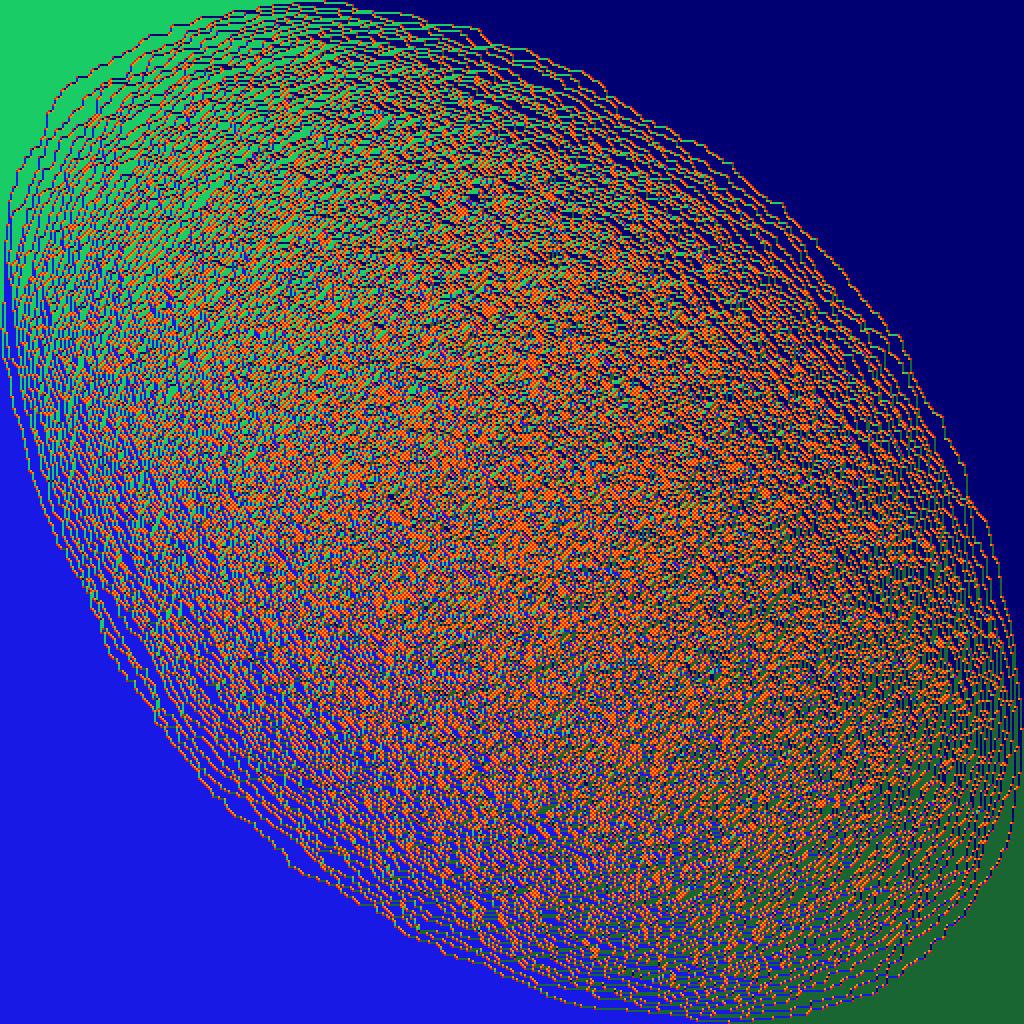}
\includegraphics[width=0.33\textwidth]{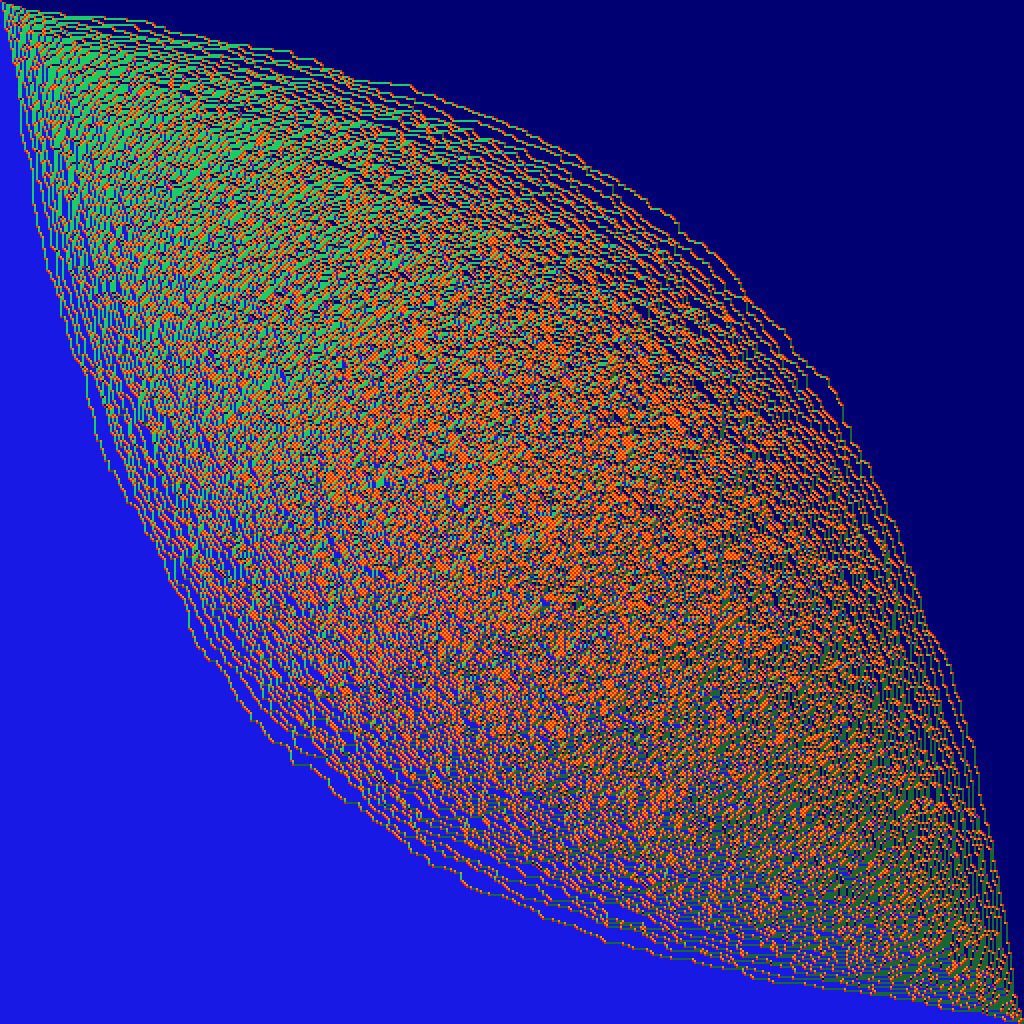}
	\caption{Typical configurations obtained using Monte Carlo.  Three cases with deformation $\lambda(x)=\pi/4+\frac{\pi \alpha}{2}(x-1/2)^2$, $\mu(y)=\frac{\pi \alpha'}{2}(y-1/2)^2$. \emph{Left:} $\alpha=\alpha'=1$. \emph{Middle:} $\alpha=0.5$, $\alpha'=-0.5$. \emph{Right:} $\alpha=1$, $\alpha'=-1$.}
	\label{fig:limitshapesfree}
\end{figure}

\paragraph{Domain wall boundary conditions with inhomogeneous weights.}
Since $\Delta=0$ here the weights can be parametrized as
\begin{align}
	\label{eq:freeweights}
	a_{jk}=\cos(\lambda_j-\mu_k) \qquad,\qquad b_{jk}=\sin(\lambda_j-\mu_k) \qquad,\qquad c=1.
\end{align}
The partition function of the inhomogeneous model is given by the deceptively simple formula  (e.g. \cite{BogoliubovPronkoZvonarev})
\begin{align}\label{eq:Zfree}
	Z_N=\prod_{1\leq j<k\leq N}\cos(\lambda_i-\lambda_j)\cos(\mu_i-\mu_j).
\end{align}
The question we want to address now is how this picture is modified in the additional presence of inhomogeneous weights \eqref{eq:freeweights}, with again slowly varying spectral parameters
\begin{align}\label{eq:scaling}
	\lambda_j= \lambda(j/N)\qquad,\qquad
	\mu_k= \mu(k/N),
\end{align}
for some reasonable functions $\lambda(x),\mu(y)$ which also ensure positive Boltzmann weights. This will be done in the next subsection with the help of a hydrodynamic theory. Before explaining this, we report in figure \ref{fig:limitshapesfree} examples of typical configurations for large $N$, where the limit shape/density profile  and the arctic curve are clearly modified.
Let us also mention that the resulting free energy $f[\lambda,\mu]=-\frac{\log Z_N}{N^2}$ can easily be evaluated from \eqref{eq:Zfree} for large $N$ as
\begin{align}\label{eq:freenrjfree}
	f[\lambda,\mu]=-\frac{1}{2}\int_0^1 dx \int_0^1 dy \log \left[\cos(\lambda(x)-\lambda(y))\cos(\mu(x)-\mu(y))\right],
\end{align}
further justifying the scaling \eqref{eq:scaling} as the most natural one to modify the limit shapes.
\subsection{Imaginary time hydrodynamic theory}
Let us consider a continuum approach, with the average vertex occupancies encoded in a density profile $\rho(x,y)$. The fact that the six vertex model can be mapped to free fermions suggests a simple hydrodynamic theory as is well known for real time evolution in the quantum setting. A fundamental role is played by the dispersion relation $\varepsilon(k)$ describing the energy of fermionic quasiparticles with momentum $k\in[-\pi,\pi]$. In the homogeneous setting the transfer matrix typically projects for large imaginary time to a ground state fermi sea $[-k_F,k_F]$, where the density is related to Fermi momentum as $\rho=\frac{k_F}{\pi}$.  As pointed out in Ref.~\cite{Abanov_hydro}, the correct hydrodynamics equation in the euclidean setting with homogeneous weights reads
\begin{align}
	\label{eq:abanov_hydro}
	\ci \partial_y z+ \partial_x \varepsilon(z)=0,
\end{align}
which is the analytic continuation (Wick rotation) $t\to -\ci y$ of the usual free fermions continuity equation $\partial_t k_F+\partial_x \varepsilon(k_F)=0$ with space dependent Fermi momentum. The interpretation of \eqref{eq:abanov_hydro} is however more subtle, since the real Fermi momentum $k_F$ is promoted to a complex number $z$, with real part being ($\pi$ times) the density, and the current determined \cite{Abanov_hydro} from the imaginary part of $\varepsilon(z)$. 

This hydrodynamic equation has been checked to provide correct predictions in numerous cases\cite{Abanov_hydro,Stephan_lectures,Bocini_2021,Pallister_2022,Abanov_2025_SciPost,ZahraDubailSchutz}. In the context of dimer coverings it is equivalent to a complex Burgers equation which was proved to provide the correct limit shape \cite{kenyon2007}. It can also be derived from transfer matrix arguments \cite{Stephan_lectures}. 
  
Adapting the above hydrodynamic description to our inhomogeneous setting is straightforward. The dispersion depends implicitly on $\lambda,\mu$ for homogeneous weights, which in our case become position dependent. Since those vary slowly in space \eqref{eq:abanov_hydro} should still hold, with meaning
\begin{align}\label{eq:mainhydro}
	\ci \partial_y z+\partial_x \varepsilon(z,\lambda(x),\mu(y))=0.
\end{align}
This is the main result we will use. As we shall see, the complications associated with the dependence on $\lambda(x),\mu(y)$ are by no means insurmountable.

For the six vertex model the dispersion is given by (e.g. \cite{zinnjustin2002,reshetikhin2010lectures})
\begin{align}
	\label{sixvertex_dispersion}
	\varepsilon(k,\lambda,\mu)=-\log \left(\frac{e^{\ci k}+\tan(\lambda-\mu)}{1-e^{\ci k}\tan(\lambda-\mu)}\right).
\end{align}
Equation \eqref{eq:mainhydro} can be solved by the method of complex characteristics, but before doing so let us rewrite it in a more symmetric form, where space $x$ and imaginary time $y$ are treated on an equal footing. Indeed the change of variables
\begin{align}\label{eq:changeofvariables}
	\tan(\omega-\lambda(x))=e^{\ci z}
\end{align}
yields
\begin{align}
	\frac{\partial_y \omega}{\sin 2(\omega-\lambda(x))}+\frac{\partial_x \omega}{\sin 2(\omega-\mu(y))}=0.
\end{align}
The characteristic lines for this last partial differential equation  can be written as
\begin{align}
	\frac{dx}{\sin 2(\omega-\mu(y))}=\frac{dy}{\sin 2(\omega-\lambda(x))}=\frac{d\omega}{0}, 
\end{align}
so the general solution is
\begin{empheq}[box=\highlightbox]{align}
\label{eq:sol}
	\int_0^y \frac{ds}{\sin 2(\omega-\mu(s))}-\int_0^x \frac{ds}{\sin 2(\omega-\lambda(s))}=G(\omega)
\end{empheq}
where $G$ is an analytic function. \eqref{eq:sol} provides a parametrization of the full set of solutions for the free inhomogeneous six vertex model, where each boundary condition corresponds to a given  $G$. We emphasize that finding the correct $G$ given the boundary conditions is often a difficult inverse problem. Several examples where this can be done for homogeneous weights are nevertheless known, often related to some form of Hilbert transform, see e.g. \cite{ADSV2016,Bocini_2021,ZahraDubailSchutz}.

For domain wall boundary conditions, we claim that the correct function $G$ is given by
\begin{align}\label{eq:Gguess}
	G(\omega)=\frac{1}{2}\int_0^1 ds\frac{\sin(\lambda(s)-\mu(s))}{\cos(\omega-\lambda(s))\cos(\omega-\mu(s))}.
\end{align} 
One can check a posteriori that this is the correct answer since it for example provides the correct boundary conditions $e^{\ci z}\in \mathbb{R}$ on all sides of the square, and reproduces the correct asymptotic form of the partition function \eqref{eq:freenrjfree}. However, this is not quite satisfactory since we did not explain how  \eqref{eq:Gguess} can be obtained in practice. This point will be clarified in section \ref{sec:arcticcurve}, the main message being that the knowledge of the arctic curve \emph{guarantees} an unambiguous determination of $G$.

To obtain the density, for a given position $(x,y)$ one has to solve \eqref{eq:sol} for $\omega$, which allows to find $e^{\ci z}$ through \eqref{eq:changeofvariables}, and finally $\rho(x,y)=\arg(e^{\ci z})$. In the frozen region $e^{\ci z}$ is a real number, or alternatively, $\omega$ is real. The arctic curve then corresponds to the case where $e^{\ci z}$ becomes complex, which generically means the hydrodynamic equation admits double roots. 

\begin{figure}[htbp]
\centering
\includegraphics[width=0.6\textwidth]{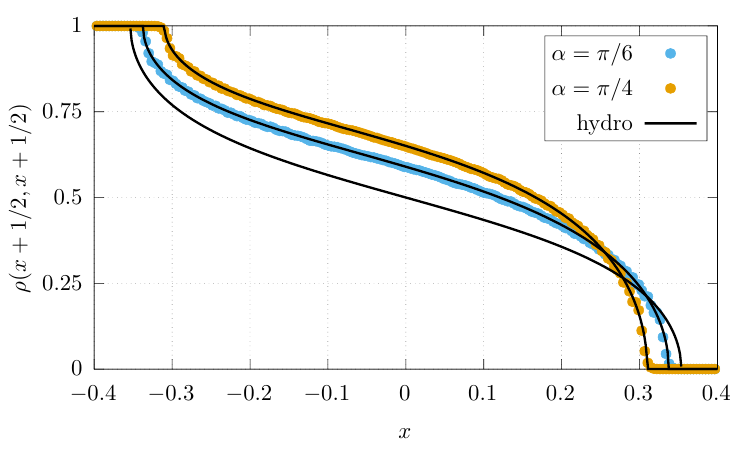}
\hfill
\includegraphics[width=0.36\textwidth]{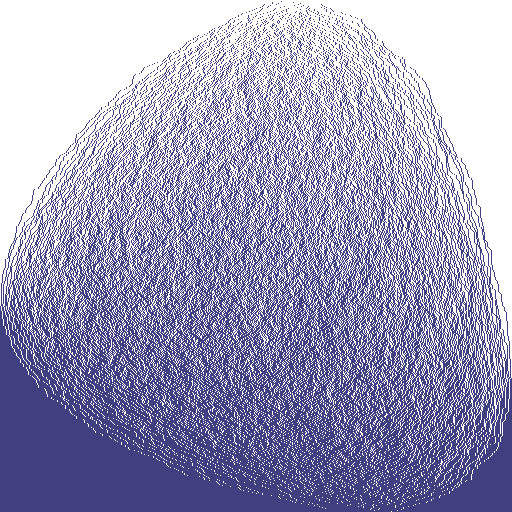}
	\caption{Left: Comparison between a numerical density profile for $N=256$ obtained using Monte Carlo and the exact formula  \eqref{eq:fulldensity} on the diagonal $y=x$ for $\lambda-\mu=\pi/4$ and two values of $\alpha=\pi/4,\pi/6$ (the homogeneous prediction $\alpha=0$ is added for reference). The hydrodynamic density corresponds to the sum of the mean occupations for vertices of type $a_1,b_2,c_1$, which is pictured on the right for $\alpha=\pi/6$. We have made numerical checks away from the diagonal as well.}
	\label{fig:densityprofile}
\end{figure}

For the simplest linear deformation
\begin{align}
	\lambda(x)&=\lambda+\alpha(x-1/2),\\
	\mu(y)&=\mu+\alpha(y-1/2),
\end{align}
all integrals can be computed exactly, and the hydrodynamic equation simplifies to
\begin{align}\label{eq:exact1}
	\frac{\sin(\mu-\alpha/2-\omega)\cos(\lambda+\alpha/2-\omega)\tan(\lambda+\alpha(x-1/2)-\omega)}{\cos(\mu+\alpha/2-\omega)\sin(\lambda-\alpha/2-\omega)\tan(\mu+\alpha(y-1/2)-\omega)}=1,
\end{align}
which ultimately leads to a quadratic equation for $e^{\ci z}$, using again \eqref{eq:changeofvariables}. The final result in the fluctuating region reads
\begin{align}
\label{eq:fulldensity}
	\rho(x+1/2,y+1/2)&=\frac{1}{\pi}\arccos \left[\frac{\sin\lambda\cos \alpha \cot(\lambda+\alpha(x-y))-\cos(\lambda+2\alpha x)}{\sqrt{\cos (2\alpha x)-\cos \alpha}\sqrt{\cos \alpha+\cos 2(\lambda+\alpha x)}}\right],
\end{align}
where we have set $\mu=0$ without loss of generality.
This exact formula compares very well with numerical simulations, as shown in figure \ref{fig:densityprofile}. We also note that the density is normalized such that $\int_0^1 dx (\rho(x,y)+y)=1$.

The exact arctic curve corresponds to asking that the argument of the $\arccos$ in \eqref{eq:fulldensity} squares to one.
For instance, for $\lambda=\pi/4$, $\mu=0$, $\alpha=\pi/4$ corresponding to the orange density curve in figure \ref{fig:densityprofile} (left), the arctic curve is given by the equation
\begin{align}
	4\cos \frac{\pi Y}{2}+4\sin \frac{\pi X}{2}-3\sin \frac{\pi(X-Y)}{2}=5.
\end{align}
 Notice that we have $\cos(\lambda(1)-\mu(0))=0$ in that case, so $\alpha=\pi/4$ is the largest deformation value available to avoid negative Boltzmann weights when $\lambda-\mu=\pi/4$. This point $(X=1,Y=0)$ also belongs to the curve. Arctic curves will be more thoroughly discussed in section \ref{sec:6vdw}.
\paragraph{Homogeneous case.} Let us briefly mention the homogeneous case, where the hydrodynamic result simplifies. The corresponding density profile reads (recall $\tan(\lambda-\mu)=b/a$):
\begin{align}
	\rho(x,y)=\frac{1}{\pi}\arccos \frac{(a/b)(y-x)+(b/a)(x+y-1)}{2\sqrt{x(1-x)}}
\end{align}
in the fluctuating region which is the interior of an ellipse  $(x-y)^2/b^2+(x+y-1)^2/a^2=1/c^2$. This result is consistent with existing literature, see e.g. \cite{ADSV2016}.
\subsection{Field theory interpretation}
The hydrodynamic equation \eqref{eq:mainhydro} can be seen as free energy minimization \cite{BloteHilhorst_1982,Nienhuis_1984,variationaldimers}, by making use of the height mapping. From this point of view vertex configurations are in a one-to-one correspondence with discrete heights $h_{m,n}$ which (say) take integer values, see figure \ref{fig:heights} below. 
In such a picture, discrete heights become continuous ones $h(x,y)$ in some scaling limit. 
\begin{figure}[htbp]
\includegraphics[width=0.33\textwidth]{figp_32_D0_tx0_ty0_2.png}
\includegraphics[width=0.33\textwidth]{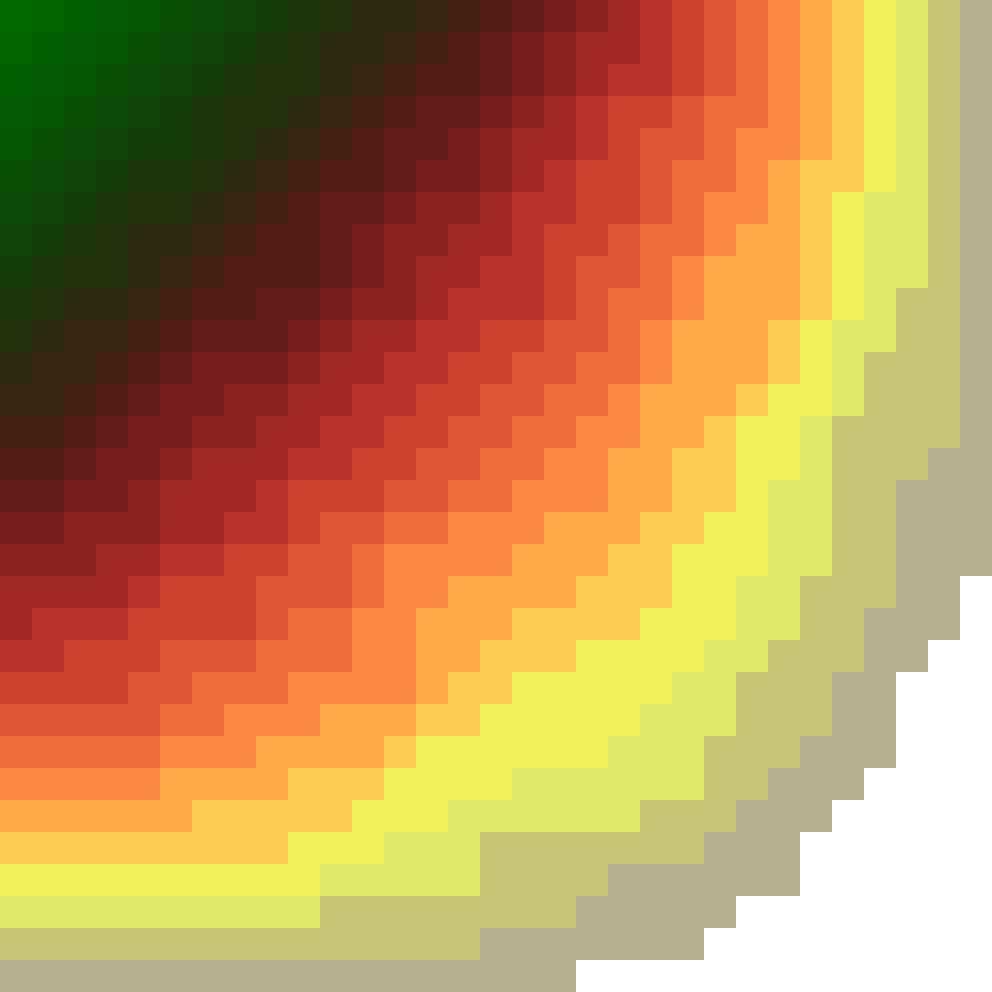}
\includegraphics[width=0.33\textwidth]{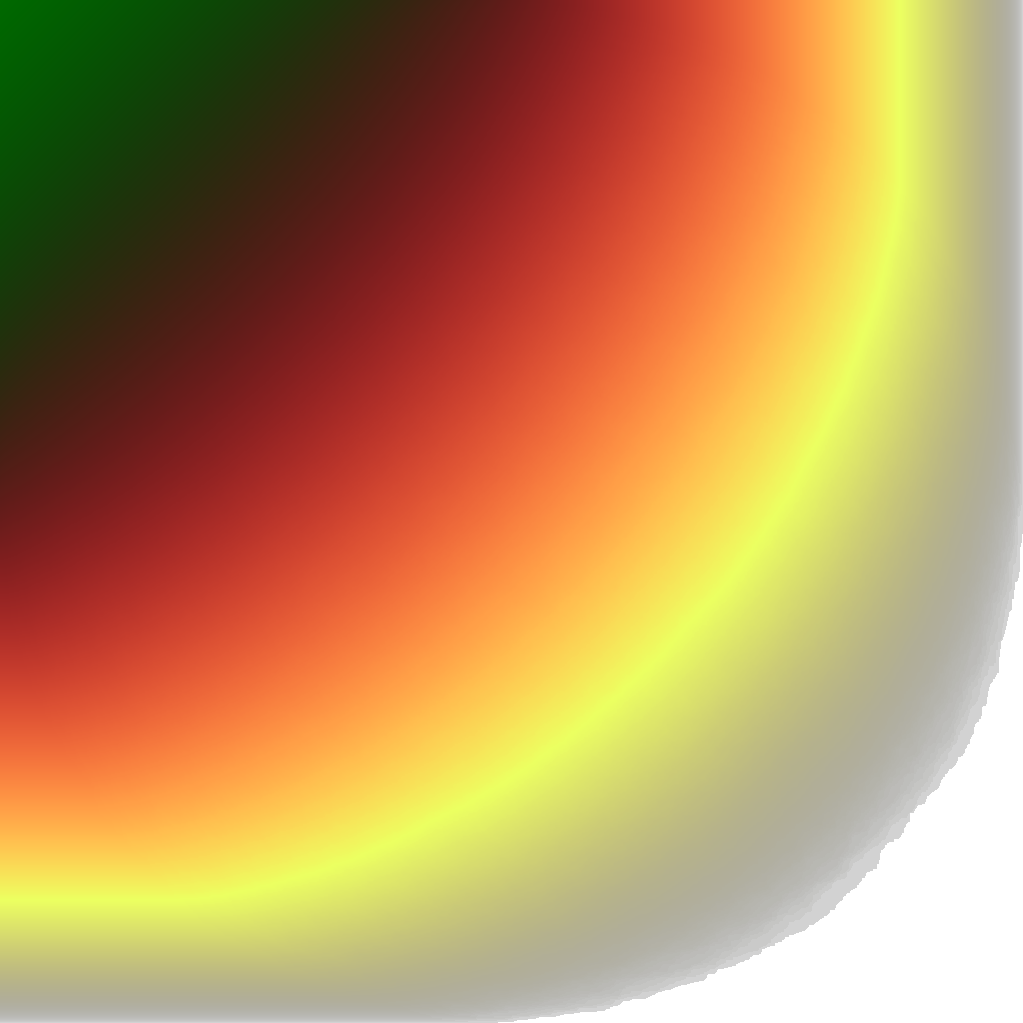}
\caption{Left: path configuration for $N=32$. Middle: corresponding height configuration, where the heights $h_{m,n}$ are level lines for the paths and are shown with different colors. Right: height configuration for $N=1024$.}
\label{fig:heights}
\end{figure} 

The limit shape is obtained by minimizing the action
\begin{align}\label{eq:actionminimization}
	\mathcal{S}[h]=\int_D dx dy \mathcal{L}(\partial_x h,\partial_y h)
\end{align}
in the fluctuating region $D$, over all heights configurations. The Lagrangian $\mathcal{L}(r,s)$ as a function of the slopes $r,s$ can be obtained by computing the free energy corresponding to a homogeneous torus partition function \cite{zinnjustin2002} with imposed slopes. In a transfer matrix picture \cite{Stephan_lectures}, the slope $r$ is related to the fermion density by a simple affine transformation, so in the following we will use $k_F$ instead. For free fermionic theories with dispersion $\varepsilon(k)$, the free energy is determined from the largest eigenvalue of the transfer matrix as \cite{Stephan_lectures,Abanov_hydro}  
\begin{align}\label{eq:exactfreeenergy}
	\mathcal{L}(k_F,s)=\int_{-k_F}^{k_F} \frac{dk}{2\pi}\varepsilon(k+\ci \nu)-\frac{\nu \, s}{\pi},
\end{align}
with $\nu$ such that $\partial_\nu \mathcal{L}=0$. The connection with the hydrodynamic approach becomes clear when introducing the complex variable $z=k_F+\ci \nu$, with $s=-\textrm{Im}\, \varepsilon(z)$ from the above condition. The hydrodynamic equation \eqref{eq:mainhydro} and its complex conjugate are then simply the Euler-Lagrange equations for the problem (\ref{eq:actionminimization}), so can be justified as such even for inhomogeneous weights. 

Now that we have access to the corresponding mean height profile $h_0$, one can ask about the fluctuations around this average profile. Following e.g. \cite{curvedlightcones}, one considers small  (quantum) fluctuations $h$ around the classical solution $h_0$. Expanding to second order
\begin{align}
	\mathcal{S}[h_0+ h]-S[h_0]=\frac{1}{2} \int \left(\mathcal{L}^{20}(\partial_x h)^2+2\mathcal{L}^{11}(\partial_x h)(\partial_y h)+\mathcal{L}^{02}(\partial_y h)^2\right) dx dy,
\end{align} 
where $\mathcal{L}^{ij}=\frac{\partial^{i+j}\mathcal{L}}{\partial k_F^i \partial s^j}$. Higher-order terms are irrelevant in $2d$, so we discard them. Hence low energy fluctuations around the solution $h_0$ are naturally described by a free scalar field theory, with parameters determined from the Hessian of \eqref{eq:exactfreeenergy}. Interpreting the inverse of the Hessian as a metric tensor $g=(\nabla^2 \mathcal{L})^{-1}$, we find using the results of \cite{Abanov_hydro}
\begin{align}\label{eq:gmunu}
	g=\pi\left(
	\begin{array}{cc}
	\frac{2}{v(z)+v(\bar{z})}	&\frac{v(z)-v(\bar{z})}{i(v(z)+v(\bar{z})} \\
	\frac{v(z)-v(\bar{z})}{i(v(z)+v(\bar{z})}	& \frac{2 |v(z)|^2}{v(z)+v(\bar{z})},
	\end{array}
	\right)
\end{align}
where $v(z)=\varepsilon'(z)$ is the analytic continuation of the Fermi speed. The determinant of the Hessian is related to the Luttinger parameter as $K=1/\sqrt{\pi^2\det(\nabla^2 \mathcal{L})}$. We have $\det \nabla^2 \mathcal{L}=1/\pi^2$ as easily follows from \eqref{eq:gmunu}, so $K=1$ here. In fact $K=1$ is a general property of free fermions theories; for interacting theories $K$ generically depends on position $K(x,y)$, see \cite{BrunDubail,Granet_2019}. 

The free field action for the fluctuations can be rewritten as
\begin{align}
	\mathcal{S}[h+h_0]-\mathcal{S}[h_0]=\frac{1}{2\pi} \int dx dy \sqrt{g} g^{\mu\nu}(\partial_\mu h)(\partial_\nu h),
\end{align}
with surface element $ds^2=g_{\mu\nu}dx^\mu dx^\nu$.
We note that this action is invariant under Weyl transformations which bring the metric $g$ to $g'=e^{2\sigma(x,y)}g$, so really only depends on the conformal class of $g$ under this identification.

A nice feature of the hydrodynamic approach is that the solution $z$ to the hydrodynamic equation $\ci \partial_y z+\partial_x \varepsilon(z,\lambda(x),\mu(y))=0$ automatically provides a set of isothermal coordinates. Indeed one can check that the action can be rewritten as
\begin{align}
	\mathcal{S}[h+h_0]-\mathcal{S}[h_0]=\frac{2}{\pi}\int (\partial_z h)(\partial_{\bar{z}}h) dz d\bar{z}
\end{align}
where $z$ solves \eqref{eq:sol},\eqref{eq:Gguess},\eqref{eq:changeofvariables}. We have $0\leq \textrm{Re}\,z\leq \pi$ and arbitrary $\textrm{Im}(z)$ so this is a boundary CFT in a slab geometry in isothermal coordinates.

Comparing with the Ising model of section \ref{sec:Ising}, we achieved a similar identification of a curved space CFT. Two main differences are that the corresponding CFT is not the same (the central charge is one for the six vertex model), and the more complicated nature of the isothermal coordinates as well as their determination. We note that this identification of the CFT would hold also for different choices of function $G(\omega)$ compared to \eqref{eq:Gguess}, which would correspond to different boundary conditions for the six vertex model. 
\pagebreak
\section[Interacting inhomogeneous six vertex model with domain wall boundaries]{Interacting six vertex model with domain wall boundaries}
\label{sec:6vdw}
In this section, we discuss the same model as in section \ref{sec:6vdwfree}, but away from the free fermion point $\Delta=0$ (or $\gamma=\pi/2$). This problem is more difficult, and in fact only the arctic curve, not the limit shape, is known even for homogeneous weights. Subsections \ref{sec:partitionfunction} and \ref{sec:polarization} are devoted to exact results and corresponding asymptotic behaviors for the partition function and boundary probabilities. We then use those asymptotic results to determine the arctic curve of the model with inhomogeneous weights in subsection \ref{sec:arcticcurve}, before checking those numerically in subsection \ref{sec:numerics}. 
\subsection{Partition function}
\label{sec:partitionfunction}
The inhomogeneous partition function $Z_N^\gamma$ was first considered in \cite{Korepin1982}, which derived recursion relations which uniquely determine it. It was then computed in determinant form in \cite{Izergin1987} (see also \cite{IzerginCokerKorepin1992}):
\begin{align}
	Z_N^\gamma&=\frac{\prod_{j,k=1}^N \sin(\lambda_j-\mu_k)\sin(\lambda_j-\mu_k+\gamma)}{\prod_{1\leq j<k\leq N}\sin(\lambda_j-\lambda_k)\sin(\mu_j-\mu_k)}\, D_N^\gamma.
\end{align}
In the previous equation $D_N^\gamma$ is the Izergin-Korepin determinant
\begin{align}\label{eq:ikdet}
	D_N^{\gamma}=\det_{1\leq j,k\leq N}\left(\int_{\mathbb{R}}dx e^{-(\lambda_j-\mu_k)x}\frac{1-e^{-\gamma x}}{1-e^{-\pi x}}\right).
\end{align}
For $\gamma=\pi/2$ this is a Cauchy determinant which completely factorizes, leading to \eqref{eq:Zfree}. 
We wish to determine the free energy $f^{\gamma}[\lambda,\mu]=-\frac{1}{N^2}\log Z_N^\gamma$ under the scaling \eqref{eq:scaling} for general $\gamma\in (0,\pi)$, generalizing the result of \cite{Korepin_2000,ZinnJustin_2000} for homogeneous weights. To gain some insights into the problem, let us first rewrite the determinant as a double sum over permutations,
\begin{align}\label{eq:proddet}
	D_N^\gamma=\frac{1}{N!}\int_{\mathbb{R}^N}dx_1\ldots dx_N \det_{1\leq j,k\leq N}(e^{-\lambda_k x_j})\det_{1\leq j,k\leq N}(e^{\mu_k x_j})\prod_{j=1}^N \frac{1-e^{-\gamma x_j}}{1-e^{-\pi x_j}},
\end{align}
and temporarily consider the specific case of linear (Toeplitz) deformations  $\lambda(x)=\lambda+\alpha(x-1/2)$, $\mu(y)=\mu+\alpha(y-1/2)$, in which case the two (Vandermonde) determinants can be computed exactly. Performing an additional rescaling $x_j\to N x_j$ of the integration variables yields
\begin{align}\label{eq:coulombgasintegral}
	D_N^\gamma=\frac{N^N}{N!}\int_{\mathbb{R}^N}dx_1\ldots dx_N 
	\prod_{1\leq j<k\leq N} 4\sinh^2 \frac{\alpha(x_j-x_k)}{2} \prod_{j=1}^N e^{-N(\lambda-\mu)x_j} \frac{1-e^{-\gamma N x_j}}{1-e^{-\pi N x_j}}.
\end{align}
This is a standard $2d$ Coulomb gas/matrix integral. Interpreting the $x_j$ as positions of particles on the real line (which have nothing to do with the particles in the underlying six vertex model), there is a competition between the repulsion generated by the $\sinh^2$ term, and a confining potential $\prod_{j=1}^N e^{-V(x_j)}$. Both terms are of the same order thanks to the rescaling, and typically particles condense on some density profile, with entropic effects negligible at the leading order. This profile can be obtained exactly, allowing an exact determination of the free energy for linear deformations, see appendix \ref{app:cg}. 

For our purposes the crucial point is that the potential can be approximated for large $N$ as
\begin{align}
	e^{-V(x_j)} \simeq e^{-N\left[(\lambda-\mu)x_j+(\gamma-\pi)x_j \Theta(-x_j)\right]} \simeq \frac{e^{-N (\lambda-\mu)x_j}}{1+e^{-N(\pi-\gamma)x_j}} 
\end{align}
without affecting the leading asymptotic behavior. In the previous equation $\Theta$ denotes the Heaviside step function. Write
\begin{align}
	\delta=\frac{\pi}{\pi-\gamma}
\end{align}
as an alternative way to parametrize interactions (the free fermion point corresponds to $\delta=2$). Upon simple rescaling and undoing all the steps starting from \eqref{eq:ikdet} yields
\begin{align}
	D_N^\gamma [\{\lambda_j\},\{\mu_k\}]\simeq \left(\frac{\delta}{2}\right)^N\det_{1\leq j,k\leq N}\left(\int_{\mathbb{R}}dx \frac{e^{-\frac{\delta}{2}(\lambda_j-\mu_k)x}}{1+e^{-\pi x/2}}\right). 
\end{align}
Said differently
\begin{align}\label{eq:nice}
	D_N^{\gamma}[\{\lambda_j\},\{\mu_k\}]\simeq \left(\frac{\delta}{2}\right)^N D_N^{\pi/2}[\{\delta\lambda_j/2\},\{\delta \mu_k/2\}],
\end{align}
so the leading asymptotic behavior can easily be deduced from the free fermion result, a tremendous simplification. While \eqref{eq:nice} was derived for a specific linear form of $\lambda(x),\mu(y)$ we expect it to hold more generally under the scaling \eqref{eq:scaling}, with the product of determinants in \eqref{eq:proddet} providing the appropriate particle repulsion.

By numerically evaluating the determinant \eqref{eq:ikdet} we find that this arguably mathematically dubious assertion\footnote{Possibly, it could be better justified by considering the ratio of determinants and using the exact inverse of a Cauchy matrix.} can be confirmed to extremely good accuracy for various choices of nonlinear deformations. In fact numerics even suggest a stronger result, namely that $\log D_N^\gamma[\{\lambda_j\}\},\{\mu_k\}]$ and $\log \left[(\delta/2)^N D_N^{\pi/2}[\{\delta\lambda_j/2\},\{\delta \mu_k/2\}]\right]$ share the same asymptotic expansion up to order $\log N$. The logarithmic correction does not appear to depend on the deformation, and coincides with the one found in \cite{Bleher} in the homogeneous setting.

Summarizing our findings, the partition function behaves as
\begin{align}\label{eq:Zexpansion}
	Z_N^{\gamma}\simeq \delta^N \prod_{1\leq j<k\leq N} \frac{\sin\delta(\lambda_j-\lambda_k)\sin\delta(\mu_j-\mu_k)}{\sin (\lambda_j-\lambda_k)\sin(\mu_j-\mu_k)} \prod_{j,k=1}^N \frac{\sin(\lambda_j-\mu_k)\sin(\lambda_j-\mu_k+\gamma)}{\sin \delta(\lambda_j-\mu_k)}
\end{align}
under the scaling \eqref{eq:scaling}, 
and the free energy $f=f^{\gamma}[\lambda(x),\mu(y)]$ is given by
\begin{align}
	f=-\frac{1}{2}\int\! \log \frac{\sin \delta(\lambda(x)-\lambda(y))\sin \delta(\mu(x)-\mu(y))\sin^2(\lambda(x)-\mu(y))\sin^2(\lambda(x)-\mu(y)+\gamma)}{\sin(\lambda(x)-\lambda(y))\sin(\mu(x)-\mu(y))\sin^2 \delta(\lambda(x)-\mu(y))} dx dy
\end{align}
where the integral is on the interval $[0,1]^2$. In the homogeneous limit $\lambda(x)\to \lambda$, $\mu(y)\to 0$, we find $f=-\log \frac{\delta\sin\lambda \sin(\lambda+\gamma)}{\sin(\delta \lambda)}$ and we recover the known result of references \cite{Korepin_2000,ZinnJustin_2000}.
\subsection{Boundary probabilities}
\label{sec:polarization}
The next observable we need to compute asymptotically is a boundary probability  which encodes the distribution of the contact point $H_n$, that is the location $(j=n,k=0)$ where the lowermost path starts traveling upwards, see figures \ref{fig:6vdwbc},\ref{fig:heights} (left). This was computed in finite size in \cite{BogoliubovPronkoZvonarev}. In particular, they found the exact formula
\begin{align}
	H_n=c^2\prod_{j=1}^{n-1}b_{j1}\prod_{j>n}^N a_{j1}
	\sum_{j=1}^n \frac{\prod_{k=2}^N a_{jk}}{\sin(\lambda_n-\lambda_j+\gamma)}\prod_{k\neq j}^n \frac{\sin(\lambda_k-\lambda_j+\gamma)}{\sin(\lambda_k-\lambda_j)} \frac{Z_{N-1}[\backslash \lambda_j,\backslash \mu_1]}{Z_N},
\end{align} 
where $\backslash \lambda_j$ means a partition function on a $(N-1)\times (N-1)$ square lattice with all $\{\lambda_1,\ldots,\lambda_N\}$ kept except for $\lambda_j$, and similarly for $\backslash \mu_1$. The notations \eqref{eq:weights} were used to make the equation fit in one line.

To understand $H_n$ asymptotically, we will make use of \eqref{eq:Zexpansion} and the remark above it:
\begin{align}\nonumber
	H_n\simeq &\frac{c^2 a_{11}}{\delta}\prod_{j=1}^{n-1}b_{j1}\prod_{j>n}^N a_{j1}
	\sum_{j=1}^n \frac{1/a_{j1}}{\sin(\lambda_n-\lambda_j+\gamma)}\prod_{k\neq j}^n \frac{\sin(\lambda_k-\lambda_j+\gamma)}{\sin \delta(\lambda_k-\lambda_j)}\prod_{k>r}^N \frac{\sin(\lambda_j-\lambda_k)}{\sin\delta(\lambda_j-\lambda_k)} \\&\times \prod_{k=1}^N \frac{b_{k1}^\delta b_{jk}^\delta\sin(\mu_k-\mu_1) }{a_{k1}b_{k1}b_{jk}\sin\delta(\mu_k-\mu_1)} 
\end{align}
where $b_{jk}^\delta=\sin(\delta(\lambda_j-\mu_k))$. Now, a standard way to proceed is to rewrite the sum as a contour integral, thanks to 
\begin{align}
	\sum_{j=1}^n h(\lambda_j)\prod_{k\neq j}^n \frac{1}{\sin  \delta(\lambda_k-\lambda_j)}=-\oint \frac{d\omega \, h(\omega)}{2\ci\pi\prod_{k=1}^n \sin \delta (\lambda_k-\omega)},
\end{align}
where the counterclockwise contour encircles all $\lambda_1,\ldots,\lambda_n$ and no other potential singularities of $h(\omega)$. We obtain
\begin{align}\nonumber
	H_n\simeq& \frac{c^2 a_{11}}{\delta b_{n1}}\oint \frac{d\omega}{2\ci\pi}\frac{1/\sin(\mu_1-\omega)}{\sin(\lambda_n-\omega+\gamma)} \prod_{k=1}^n\frac{\sin(\lambda_k-\omega+\gamma)\sin(\lambda_k-\mu_1)}{\sin \delta(\lambda_k-\omega)}\prod_{k>n}^N \frac{\sin(\omega-\lambda_k)\sin(\lambda_k-\mu_1+\gamma)}{\sin \delta(\omega-\lambda_k)}\\
	&\times \prod_{k=1}^N \frac{\sin \delta(\lambda_k-\mu_1)\sin \delta(\omega-\mu_k)\sin(\mu_k-\mu_1)}{\sin(\lambda_k-\mu_1+\gamma)\sin(\lambda_k-\mu_1)\sin(\omega-\mu_k)\sin \delta(\mu_k-\mu_1)}.
\end{align}
In a large deviation regime, $H_n$ behaves as
\begin{align}
	H_{xN} \simeq e^{-N \mathcal{I}_x(\omega)}.
\end{align}
The corresponding rate function can be written as
\begin{align}\nonumber
	\mathcal{I}_x(\omega)=&\int_0^x ds \log \frac{\sin \delta(\lambda(s)-\omega)}{\sin(\lambda(s)-\omega+\gamma)\sin(\lambda(s)-\mu(0))}+\int_x^1 ds \log \frac{\sin \delta(\lambda(s)-\omega)}{\sin(\lambda(s)-\omega)\sin(\lambda(s)-\mu(0)+\gamma)}\\ \label{eq:largedeviation}
	&-\int_0^1 ds \log \frac{\sin \delta(\mu(s)-\omega) \sin \delta(\lambda(s)-\mu(0))\sin(\mu(s)-\mu(0))}{\sin (\mu(s)-\omega)\sin(\lambda(s)-\mu(0)+\gamma)\sin(\lambda(s)-\mu(0))\sin \delta(\mu(s)-\mu(0))}, 
\end{align}
where $\omega$ solves the saddle point equation
\begin{align}\label{eq:spe}
	\int_0^x ds \cot(\lambda(s)-\omega+\gamma)+\int_x^1 ds \cot(\lambda(s)-\omega)=H(\omega),
\end{align}
with 
\begin{align}\label{eq:Hfunction}
	H(\omega)=\int_0^1 ds\left[\delta \cot \delta(\lambda(u)-\omega)-\delta \cot \delta(\mu(s)-\omega)+\cot(\mu(s)-\omega)\right].
\end{align}
Equations \eqref{eq:largedeviation},\eqref{eq:spe},\eqref{eq:Hfunction} are the main technical results we will need to determine the arctic curve. The on-shell derivative
\begin{align}\label{eq:onshellderivative}
	\frac{d\mathcal{I}_x(\omega)}{dx}=\log \frac{\sin (\lambda(x)-\omega)\sin(\lambda(x)-\mu(0)+\gamma)}{\sin(\lambda(x)-\omega+\gamma)\sin(\lambda(x)-\mu(0))}
\end{align}
will also turn out to be useful. 
\subsection{Determination of the arctic curve}
\label{sec:arcticcurve}
\begin{figure}[htbp]
\includegraphics[width=0.33\textwidth]{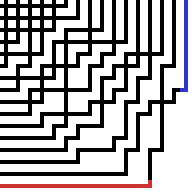}
\includegraphics[width=0.33\textwidth]{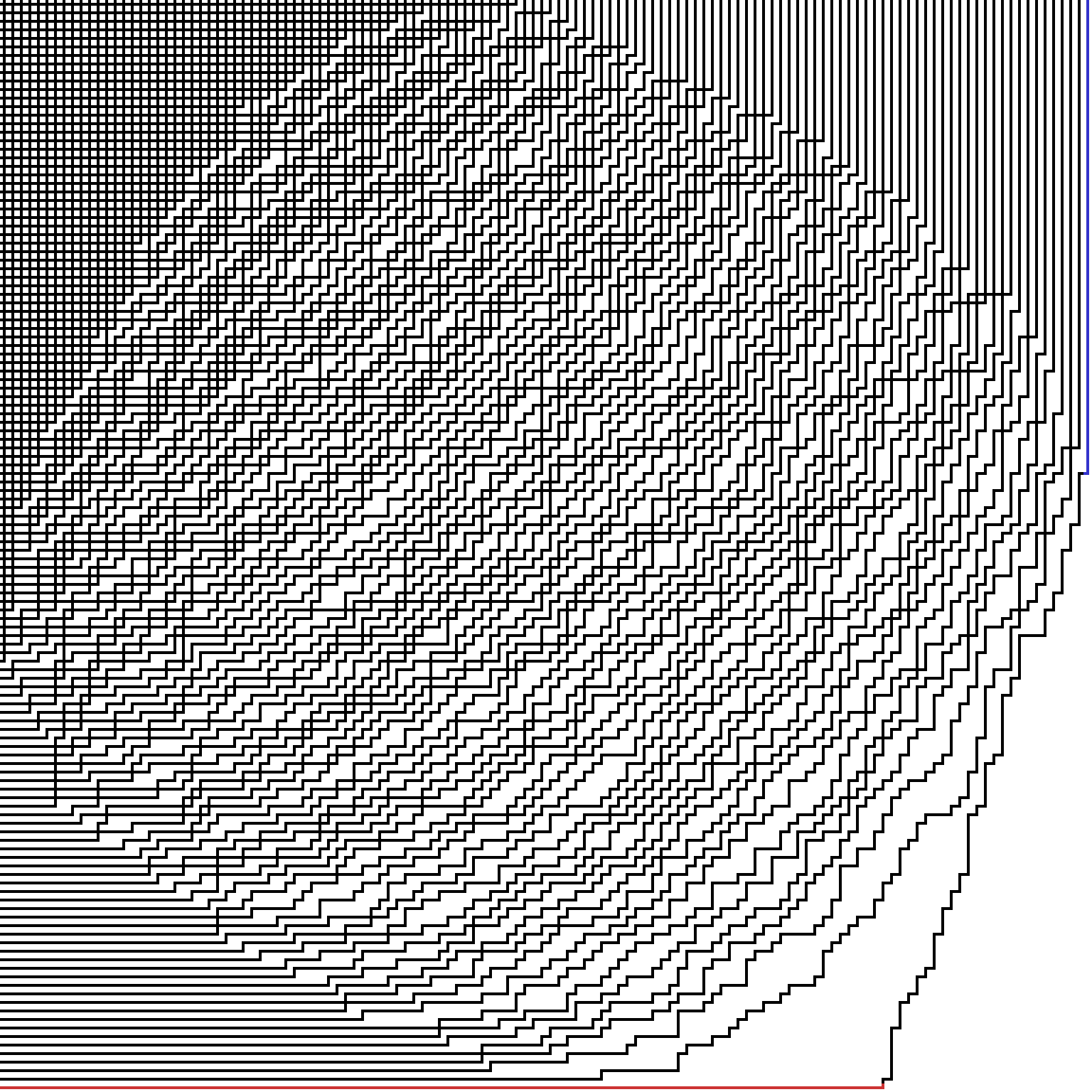}
\includegraphics[width=0.33\textwidth]{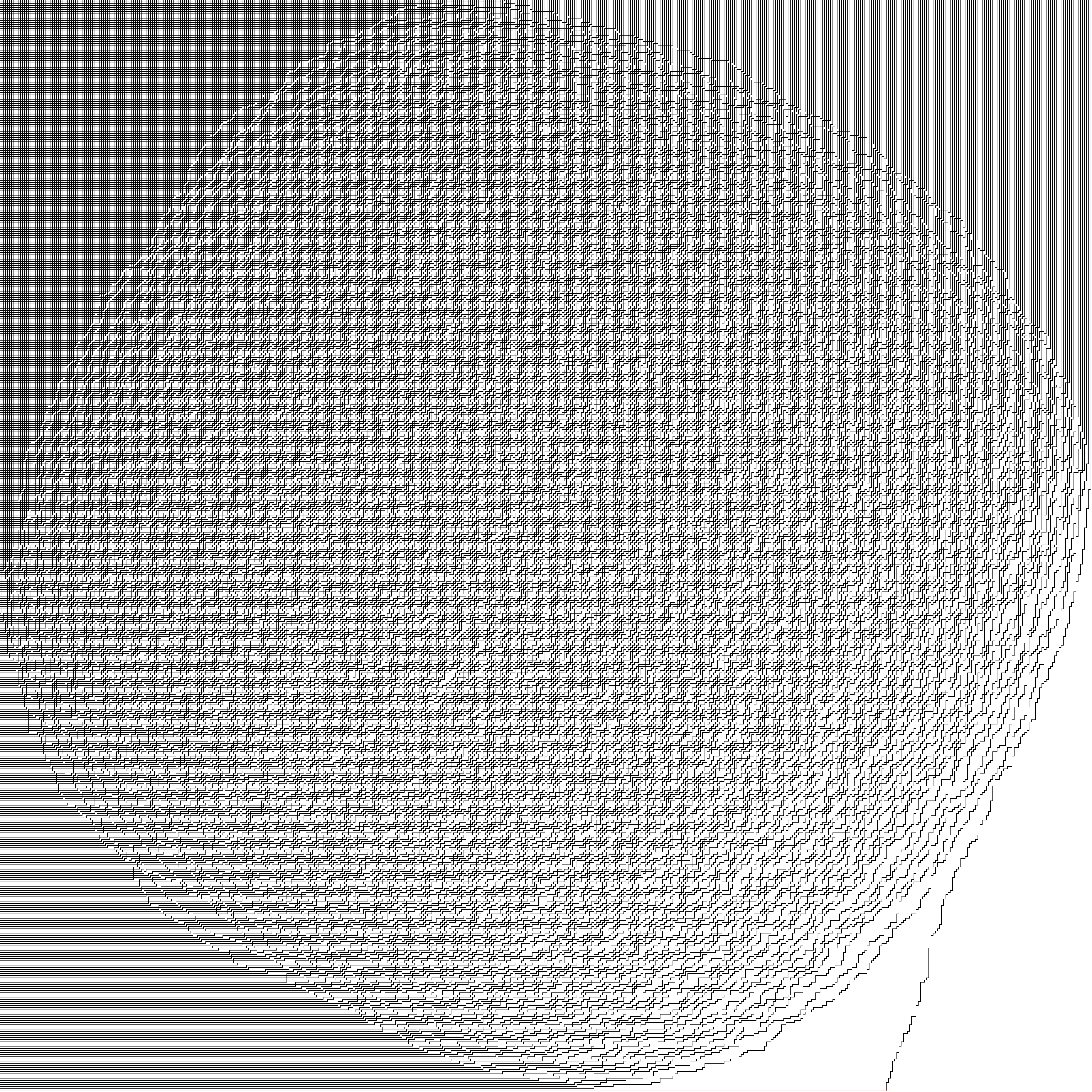}
\caption{Illustration of the tangent method on typical path  configurations. The red part is imposed, but not the rest of the trajectory. Left: $N=16$. Middle: $N=128$. Right: $N=512$. The relevant path follows a geodesic before meeting (tangentially) the arctic curve, and following it until the right boundary is reached. The blue part corresponds to the trivial upwards trajectory once the right boundary is reached.}
	\label{fig:tangent}
\end{figure}

We will determine the arctic curve using the tangent method \cite{TangentMethod}. To better explain this approach, let us first  determine the contact point $(X_c,y=0)$ of the arctic curve with the lower boundary, which directly relates to our previous computation.
\paragraph{Contact point.}
Observe first that $\mathcal{I}_x(\omega)=-\lim_{N\to\infty }\frac{\log H_{xN}}{N}$ happens to vanish for $\omega=\mu(0)$, see \eqref{eq:largedeviation}. Since $\sum_{n=1}^N H_n=1$, the saddle point $\omega=\mu(0)$ simply corresponds to the contact point $X_c$ in the scaling limit. Making use of the saddle point equation \eqref{eq:spe}, it is given by
\begin{align}
	\int_0^{X_c} ds \cot(\lambda(s)-\mu(0)+\gamma)+\int_{X_c}^1 ds \cot(\lambda(s)-\mu(0))=H(\mu(0)).
\end{align}
\paragraph{Geodesic tangent method.}
The basic idea of the tangent method \cite{TangentMethod} is to impose a different location for the point $x(\omega)>X_c$ where the lower rightmost path starts traveling upwards, see figure \ref{fig:tangent} where this corresponds to the end of the red portion of the path. The free energy cost associated to this is precisely given by $\mathcal{I}_x(\omega)$. Then, the method can be formulated as a principle of least action for the most likely trajectory $(\mathbf{x}(t),\mathbf{y}(t))$, where the underlying Lagrangian can be understood, as we shall see, from purely free particle arguments. In the following we will parametrize the arctic curve as $(X(\omega),Y(\omega))$, where we have already established that $X(\mu(0))=X_c$, $Y(\mu(0))=0$.

Let us now explain how this idea can be used in practice. In the scaling limit, the contribution of the most likely trajectory can be written as a sum of four terms
\begin{align}\label{eq:tangentmethod}
	\mathcal{I}_x(\omega)=\left(\int_{t_0}^{t_1}+\int_{t_1}^{t_2}+\int_{\omega}^{t_3}+\int_{t_3}^{t_4}\right) \mathscr{L}(\dot{\bx},\dot{\by},\lambda(\bx),\mu(\by))dt, 
\end{align}
where the first corresponds to the red part in figure \ref{fig:tangent} with $\dot{\by}=0$, and the last corresponds to the blue part with $\dot{\bx}=0$. The second term corresponds to a free trajectory until the path meets the arctic curve at $\bx(t_2)=X(\omega)$. The third term corresponds to the part which follows the arctic curve until $\bx(t_3)=1$, as can most clearly be seen in the right part of the figure.

The aim is to find the trajectory $(\bx(t),\by(t))$ which minimizes the rhs of \eqref{eq:tangentmethod}, and only at this minimum does the rhs  equal $\mathcal{I}_x(\omega)$. The exact knowledge of the integrable input $\mathcal{I}_x(\omega)$ then allows to reconstruct the arctic curve. The name tangent method comes from the fact that the path joins the arctic curve tangentially, as making an angle would have a finite free energy cost. The main difficulty compared to most studies using this method \cite{DiFrancesco_2018_aad028,Colomo_2018,Aggarwal_2019,Debin_2020,Ruelle_2022,Stephan_2022,DiFrancesco_2024} is that the trajectory corresponding to the second part is not a straight line anymore, due to the inhomogeneous weights. One has to find the proper \emph{geodesic path} first, yet another manifestation of curved space time. We note that Refs.~\cite{DiFrancescoGuitter,DiFrancescoGuitter2} previously encountered a similar situation in free fermionic models, where the correct geodesic path also had to be determined. The addition of interactions in the present case is not expected to question the validity of the approach. 

\begin{figure}[ht!]
\includegraphics[width=0.33\textwidth]{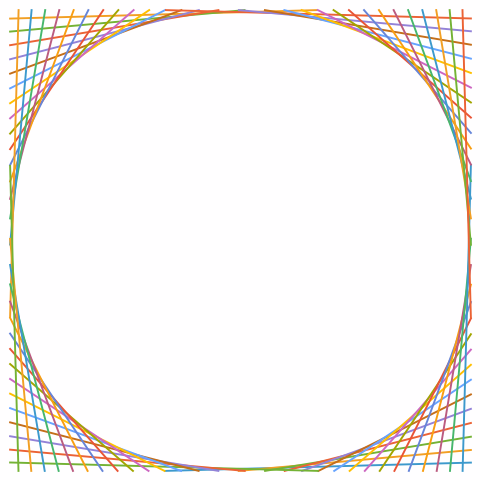}
\includegraphics[width=0.33\textwidth]{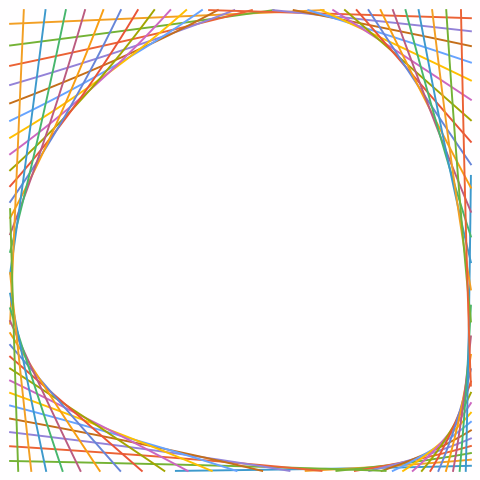}
\includegraphics[width=0.33\textwidth]{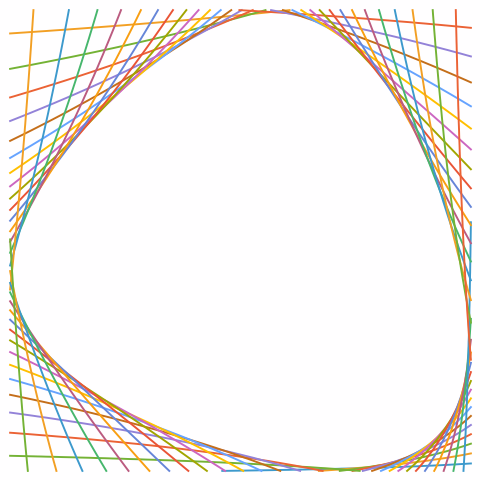}\\
\includegraphics[width=0.33\textwidth]{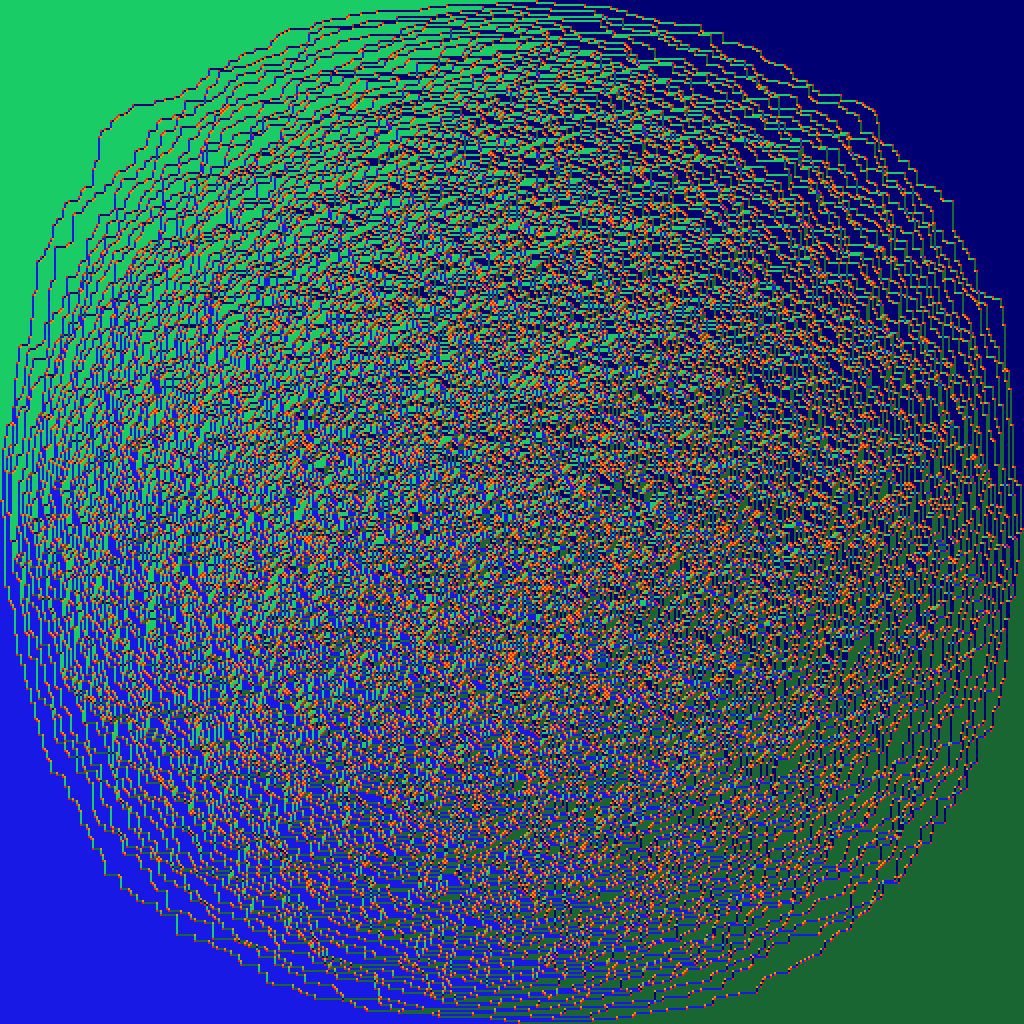}
\includegraphics[width=0.33\textwidth]{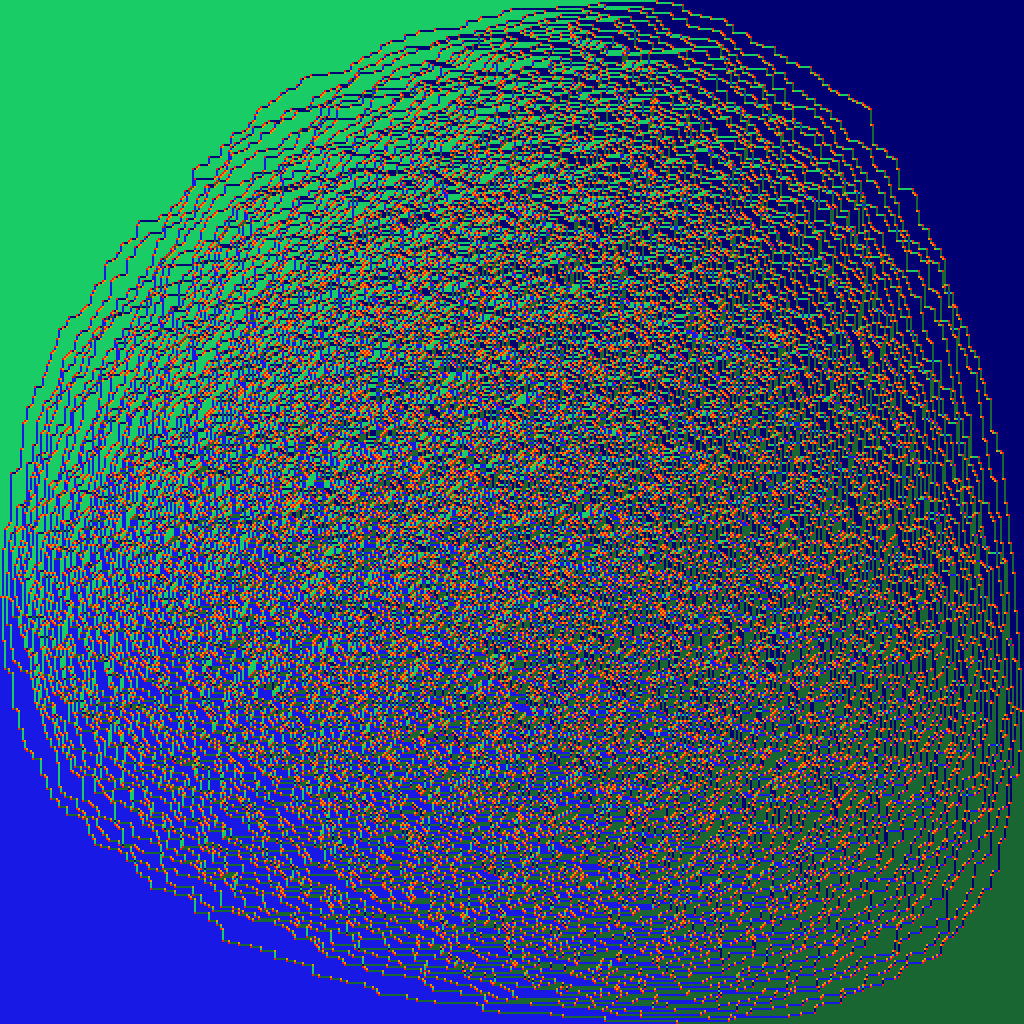}
\includegraphics[width=0.33\textwidth]{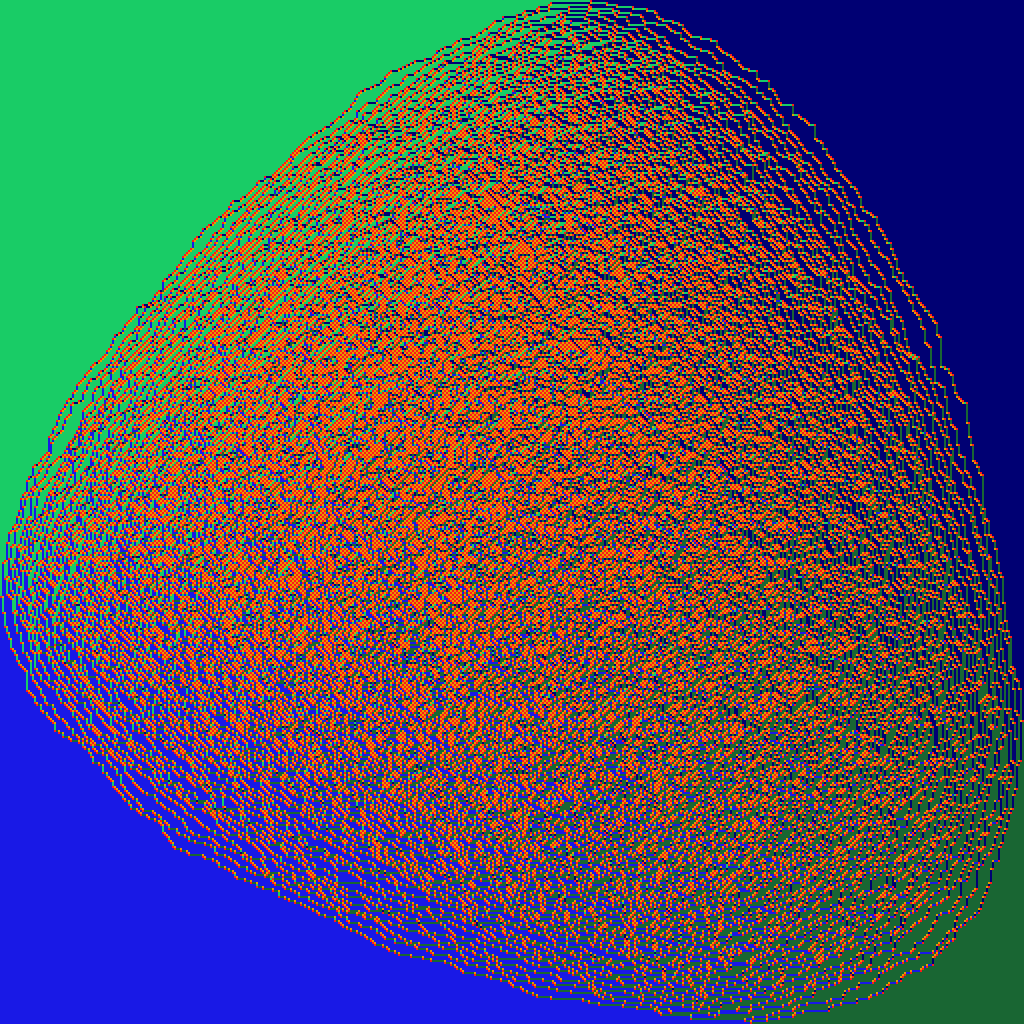}
\caption{Top figures: arctic curves of the six vertex model as caustics for the curved rays \eqref{eq:rays} and the choice of inhomogeneous deformation \eqref{eq:lineardeformationinteracting}. Left: $\Delta=0.8$, $\lambda=\frac{\pi-\gamma}{2}$, $\alpha=0$ (homogeneous weights, with no curvature effects, for comparison). Middle: $\Delta=0.8$, $\lambda=\frac{\pi-\gamma}{2}$, $\alpha=\pi/5$. Right: $\Delta=-0.6$, $\lambda=\frac{\pi-\gamma}{2}$, $\alpha=\pi/10$. Bottom figures: corresponding typical configurations for $N=512$, with good agreement.}
\label{fig:ac_interactions}
\end{figure}

\begin{figure}[htbp]
\includegraphics[width=0.33\textwidth]{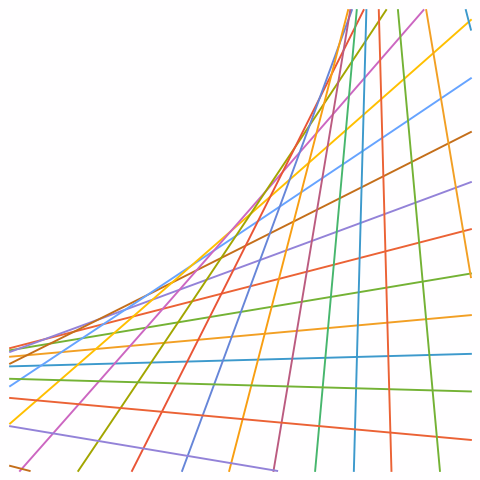}\hfill
\includegraphics[width=0.33\textwidth]{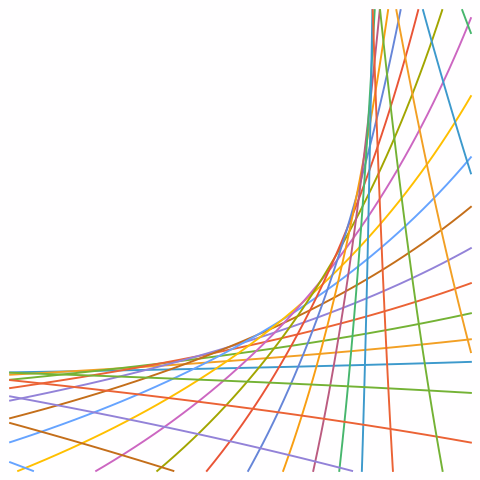}\hfill
\includegraphics[width=0.33\textwidth]{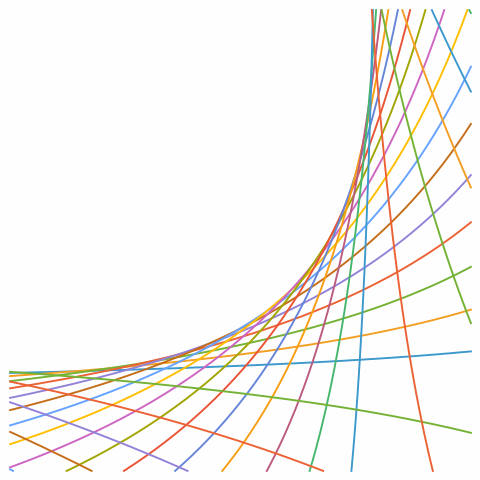}
	\caption{Rays $R(\omega)=0$ near $x=1,y=0$ for the exact same parameter values as in figure \ref{fig:ac_interactions}. The left figure displays homogeneous parameters, where the rays are straight lines, despite the caustic being a non trivial curve. The other two figures display curved rays, similar to those found in  \cite{DiFrancescoGuitter} in a different model.}
	\label{fig:curvature}
\end{figure}

Let us now proceed with this geodesic computation. The exact expression \cite{TangentMethod} for the homogeneous Lagrangian $\mathscr{L}$ for one path in the vacuum is
\begin{align}\label{eq:freeratefunction}
	\mathscr{L}(u,v,\lambda,\mu)=u\log \frac{u-r_s}{u}+v\log \frac{v-r_s}{v}-(u+v)\log \frac{\sin(\lambda-\mu)}{\sin(\lambda-\mu+\gamma)}
\end{align}
where $r_s$ solves the equation $(u-r)(v-r)=r^2 \frac{\sin^2 \lambda}{\sin^2\gamma}$. By separation of scales, the inhomogeneous Lagrangian is simply obtained by promoting $\lambda,\mu$ to functions $\lambda(x),\mu(y)$ as in the previous section. The equations of motion for the geodesic path in the interval $[t_1,t_2]$ read
\begin{align}
	\lambda'(\bx)\frac{\partial \mathscr{L}}{\partial \lambda(\bx)}&=\frac{d}{dt}\frac{\partial \mathscr{L}}{\partial\dot{\bx}},\\
	\mu'(\by)\frac{\partial \mathscr{L}}{\partial \mu(\by)}&=\frac{d}{dt}\frac{\partial \mathscr{L}}{\partial\dot{\by}}.
\end{align}
To proceed further, we consider the variation of \eqref{eq:tangentmethod} as $\omega$ is changed. 
The most complicated contribution is that of the second term. 
Taking derivative, we find
\begin{align}
	\frac{d}{d\omega}\int_{t_1}^{t_2}&=\left[\frac{d\bx}{d\omega}\frac{\partial \mathscr{L}}{\partial \dot{\bx}}+\frac{d\by}{d\omega}\frac{\partial \mathscr{L}}{\partial \dot{\by}}\right]_{t_1}^{t_2}\\
	&=\mathscr{L}(\dot{X},\dot{Y})-\frac{dx}{d\omega}\mathscr{L}^{(1,0)}(\dot{\bx}(t_1),\dot{\by}(t_1)),
\end{align}
using the equations of motion and performing integration by parts.  \eqref{eq:tangentmethod} and \eqref{eq:onshellderivative} imply
\begin{align}
	\log \frac{\sin (\lambda(\bx)-\omega)\sin(\lambda(\bx)-\mu(0)+\gamma)}{\sin(\lambda(\bx)-\omega+\gamma)\sin(\lambda(\bx)-\mu(0))}=-\log \frac{\sin(\lambda(\bx)-\mu(0))}{\sin(\lambda(\bx)-\mu(0)+\gamma)}-\mathscr{L}^{(10)}(\dot{\bx}(t_1),\dot{\by}(t_1)).
\end{align}
with $\mathscr{L}^{(10)}(u,v)=\partial_u \mathscr{L}(u,v)$. 
Using \eqref{eq:freeratefunction} we obtain the slope at $(x(\omega),0)=(\bx(t_1),\by(t_1))$:
\begin{align}
	\frac{\dot{\by}(t_1)}{\dot{\bx}(t_1)}=\frac{\sin(\mu(0)-\omega) \sin(\mu(0)-\omega+\gamma)}{\sin(\lambda(\bx)-\omega) \sin(\lambda(\bx)-\omega+\gamma)},
\end{align}
and this implies the full geodesic path satisfies
\begin{align}
	\frac{\dot{\by}(t)}{\dot{\bx}(t)}=\frac{\sin(\mu(\by)-\omega) \sin(\mu(\by)-\omega+\gamma)}{\sin(\lambda(\bx)-\omega) \sin(\lambda(\bx)-\omega+\gamma)}.
\end{align}
This means, going back to to labeling space coordinates as $x,y$, instead of $\bx(t),\by(t)$:
\begin{align}
	\int_{x(\omega)}^{x} \frac{du}{\sin(\lambda(u)-\omega)\sin(\lambda(u)-\omega+\gamma)}=\int_0^{y} \frac{dv}{\sin(\mu(v)-\omega)\sin(\mu(v)-\omega+\gamma)},
\end{align}
where $x=x(\omega)$ solves the saddle point equation \eqref{eq:spe}. Said differently $R(\omega)=0$, where
\begin{empheq}[box=\highlightbox]{align}
	\nonumber
	R(\omega)=&\int_0^{y}\!\frac{ds\,\sin \gamma}{\sin(\mu(s)-\omega) \sin(\mu(s)-\omega+\gamma)}-\int_0^{x}\! \frac{ds\,\sin \gamma}{\sin(\lambda(s)-\omega)\sin(\lambda(s)-\omega+\gamma)}\\
	\label{eq:rays}&-\int_0^1 ds \left[\frac{\sin(\lambda(s)-\mu(s))}{\sin(\lambda(s)-\omega)\sin(\mu(s)-\omega)}-\frac{\delta\sin \delta(\lambda(s)-\mu(s))}{\sin \delta(\lambda(s)-\omega)\sin \delta(\mu(s)-\omega)}\right].
\end{empheq}
Equation  
\eqref{eq:rays} is the central result of this section, generalizing that of \cite{colomo2010arctic} to inhomogeneous weights $\lambda(x),\mu(y)$. The equation $R(\omega)=0$ defines a family of curved rays (the geodesics) in the $(x,y)$ plane indexed by $\omega\in[\lambda(1)+\gamma-\pi,\mu(0)]$, which fill the entire bottom right frozen region (the others can be deduced by symmetry). The arctic curve is then the boundary of this frozen region, where the solution to the equation $R(\omega)=0$ becomes complex. Hence the arctic curve is the caustic for these rays, and can be determined from the two conditions
\begin{align}\label{eq:caustic}
	R(\omega)=0 \qquad,\qquad \frac{d R(\omega)}{d\omega}=0.
\end{align}
Figure \ref{fig:ac_interactions} shows the rays and the caustics for
the choice of linear deformation
\begin{align}\label{eq:lineardeformationinteracting}
	\lambda(x)=\frac{\pi-\gamma}{2}+\alpha(x-1/2) \qquad,\qquad
	\mu(y)=\alpha(y-1/2),
\end{align}
together with typical configurations, with excellent visual agreement.
We note that the arctic curve can be obtained in closed form, however the exact expression is not particularly illuminating. 
Let us nevertheless report the formula for the contact point, which reads
\begin{align}
	x_c&=\frac{1}{\alpha}\arctan \left[\frac{\sin\alpha+\cot \frac{\gamma}{2}\left(\cos \alpha-\delta \cot \delta \alpha \sin\alpha\right)}{\cos \alpha+\sin\alpha(\tan \frac{\gamma}{2}+\delta \cot \delta \alpha)}\right].
\end{align}
To highlight curvature effects, we show in figure \ref{fig:curvature} a zoom corresponding to the region near $x=1,y=0$ in figure \ref{fig:ac_interactions}.
\paragraph{Relation to free fermions hydrodynamics.} At the free fermion point our main equation $R(\omega)=0$ can be rewritten as 
\begin{align}\label{eq:raysfree}
	\int_0^y \frac{ds}{\sin 2(\omega-\mu(s))}-\int_0^x \frac{ds}{\sin 2(\omega-\lambda(s))}=\frac{1}{2}\int_0^1 ds \frac{\sin(\lambda(s)-\mu(s))}{\cos(\lambda(s)-\omega)\cos(\mu(s)-\omega)}.
\end{align}
This coincides exactly with the hydrodynamic solution  \eqref{eq:sol}, with $G(\omega)$ given by \eqref{eq:Gguess}. The result \eqref{eq:raysfree} thus provides a direct way to find the correct hydrodynamic function $G(\omega)$. Indeed, recall that the hydrodynamic complex Fermi momentum $z=z(x,y)$ is related to $\omega$ by $\tan(\omega-\lambda(x))=e^{\ci z}$, with density $\rho(x,y)= \textrm{Re}(z)/\pi$. The case of non trivial density $0<\rho(x,y)<1$ corresponds to complex $\omega$, and real $\omega$ corresponds to frozen densities. 

The equation for the curved rays $R(\omega)$, written in the right parametrization, already contains \emph{full information} regarding the limit shape/density profile in the fluctuating region, not only the arctic curve. This assertion can be seen as a consequence of the consistency of the tangent method with the hydrodynamic approach.
\subsection{Numerical verifications}
\label{sec:numerics}
As was shown in figure \ref{fig:ac_interactions}, typical configurations for large $N$ seem to match reasonably well our prediction for the arctic curves. We provide in this subsection more quantitative verifications that \eqref{eq:rays} is indeed correct.
To identify the arctic curve, we consider the trajectory of the lower rightmost path and compute numerically its statistics using Monte Carlo. In particular the mean allows to reconstruct a portion of the curve, with the other portions obtained by similar methods.

\begin{figure}[htbp]
\includegraphics[width=0.48\textwidth]{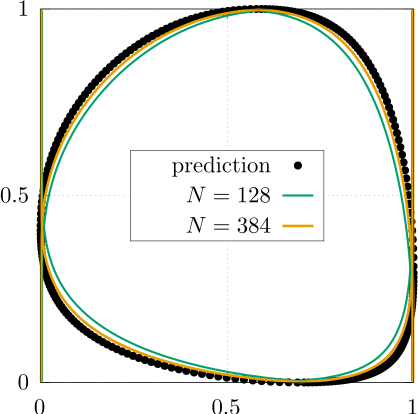}
\hfill
\includegraphics[width=0.48\textwidth]{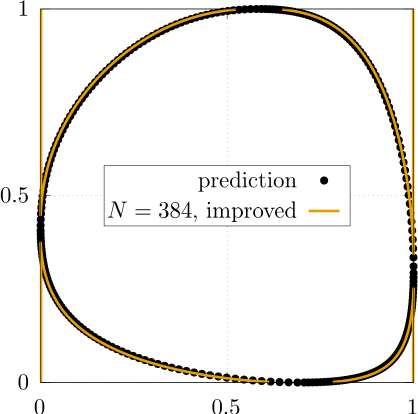}
\caption{Comparison between a numerical determination of the arctic curve and the prediction \eqref{eq:rays} for $\Delta=0.8$, $\alpha=\pi/5$. \emph{Left:} Raw data for $N=128,384$. \emph{Right:} Improved data, based on Tracy-Widom scaling for the edge distribution, as discussed in the text. The agreement is already excellent for $N=384$.}
\label{fig:ac_numerics}
\end{figure}

Figure \ref{fig:ac_numerics} (left) shows a comparison between finite size results for $N=128,384$ and the predicted arctic curves from \eqref{eq:rays}, \eqref{eq:caustic}. As can be seen the agreement is fair, and even though it improves for large $N$ finite-size effects are still noticeable. 

To improve the agreement, we make use of the Pokrovsky-Talapov \cite{PokrovskyTalapov}, KPZ \cite{KPZ_1986} scaling $N^{1/3}$ near the arctic curve and the fact that on this scale the distribution of the exterior particle follows the Tracy-Widom distribution \cite{TracyWidom_tw}. This is well established in the free fermionic setting (see \cite{Joh_review} for a review), and has been checked numerically in the presence of interactions with homogeneous weights \cite{Lyberg_2023,Prahofer_2023}. We do not expect inhomogeneous weights to affect this behavior (see e.g. \cite{Stephan_edge} for a simple dilution argument why that is). For this (asymmetric) distribution the ratio between the mean and the standard deviation is $\simeq -1.964$, so shifting the exterior particle distribution by $\pm 1.964$ times its standard deviation effectively removes the leading finite size correction. As can be seen in figure \ref{fig:ac_numerics} (right), the agreement for $N=384$ improves by an order of magnitude, and becomes nearly perfect.    
\section{Conclusion}
In this paper we have studied inhomogeneous deformations of well known free and integrable lattice models of statistical mechanics, with motivations from massless field theory, integrability, and low-dimensional quantum systems. Our focus was on slowly varying clean deformations, which naturally lead to CFTs in curved space, with a space-time metric which we determined exactly. 
Our two main examples were the inhomogeneous Ising model and the (more technically involved) inhomogeneous six vertex model with domain wall boundary conditions. 

Crucial to our analysis of the Ising model was the ability to vary the underlying Fermi speed $v_F$ which rescales the low energy CFT spectrum. This is possible for lattice models where one can break isotropy while still maintaining criticality (indeed in an isotropic model both space directions are equivalent and this effectively enforces  $v_F=1$). By making anisotropy change slowly in space we found a simple lattice realization of an inhomogeneous CFT. While we have not attempted to derive this from the lattice and used high precision numerics instead, the results of \cite{Chelkak_2024,Fagotti2025} might be useful to achieve this more ambitious goal.

There are of course several similar models which we have not discussed in the paper. For example, we expect our results to readily apply to the three state Potts model, which also has a line of anisotropic critical points all described by (a sector of) the  minimal model with central charge $c=4/5$ \cite{Yellowbook,Mussardo}. Lattice computations are much more difficult, but our predictions could be confirmed by numerical simulations based on Monte Carlo or matrix product states. Our logic can also be applied to higher dimensions, where the underlying  conformal field theories are more complicated.

For the six vertex model, the deformation of spectral parameters \eqref{eq:weights} is not at all new. In fact, it lies at the root of the integrability structure. However the study of these inhomogeneous features has often been restricted to algebraic aspects (Yang Baxter relation, etc), which were seen as a technical necessity to solve the homogeneous model, study generalizations with some form of translation invariance, or implement change in boundary conditions (see, however, \cite{Keating2022}). In this paper we used a different logic: we considered as such a truly inhomogeneous weights scaling limit, \eqref{eq:scaling}, very similar to that of the inhomogeneous Ising model. Our main result in the interacting case was a determination of the arctic curve which separates the fluctuating region from ordered (frozen) ones. This is based on an exact formula for a family of \emph{curved} rays \eqref{eq:rays} which allows to reconstruct the arctic curve as a their caustic, generalizing the homogeneous result of \cite{colomo2010arctic,TangentMethod} in the disordered regime. In this formula the inhomogeneous spectral parameters appear in a very natural way, similar to a would be continuum version of the Yang-Baxter equation.

It would be interesting to investigate the other ferroelectric and antiferroelectric regime as well. There are known extra technical difficulties in the homogeneous antiferroelectric case \cite{ZinnJustin_2000,ColomoPronkoZinn} with $\Delta<-1$, even though it is not unreasonable to expect at least the case of linear deformation to admit an exact solution as well. 

We have also discussed how to understand the free case using a hydrodynamic theory put forward in \cite{Abanov_hydro}, and used it to determine the full density profile and the underlying inhomogeneous CFT. We checked some of those predictions using Monte Carlo numerical simulations. While we have not attempted to recover these results  from exact lattice calculations, this is certainly feasible, for example using the results of \cite{ColomoDiGiulioPronko,Aggarwal2023}. We note that adapting the hydrodynamic theory to include interactions still poses significant technical challenges \cite{zinnjustin2002,reshetikhin2010lectures}, despite a solution to the five vertex model \cite{deGier2021}. This is in sharp contrast with the real time case, where generalized hydrodynamics\cite{GHD1,GHD2} provides the answer for interacting integrable quantum systems, with the results of \cite{GHD3} closest in logic to the present paper. 

Back to a field theory perspective on the six vertex model with domain wall boundaries, one expects a inhomogeneous free field theory with Luttinger parameter which depends on position \cite{BrunDubail,Granet_2019}, however proper identification of the metric tensor and Luttinger parameter would require a solution to the aforementioned hydrodynamic theory, or exact input from the lattice. Finally, one could study integrability breaking deformations, or deformations of non integrable critical models, for which our field-theoretical arguments should still hold. 
\section*{Acknowledgments}
I acknowledge discussions with F. Colomo and A. Fedorenko.

\pagebreak
\begin{appendix}
\addtocontents{toc}{\protect\setcounter{tocdepth}{1}}
\numberwithin{equation}{section}
\section{Free fermions and the Ising model}
\label{app:IsingTMfermions}
In this appendix, we gather several calculations and free fermions concepts relevant to section \ref{sec:Ising}. 
Most of what can be found here is based on the review paper \cite{SchultzMattisLieb} and the Nambu formalism.
\subsection{Free fermions reminder}
\label{app:freefermions}
\paragraph{Quadratic Hamiltonians.}
Consider a Hamiltonian
\begin{align}
	H&=A_{ij}c_i^\dag c_j +\frac{1}{2}\left(B_{ij}c_i^\dag c_j^\dag-C_{ij}c_i c_j\right),
\end{align}
where the $c_i,c_j^\dag$ are fermionic operators obeying the anticommutation relations $\{c_i,c_j^\dag\}=\delta_{ij}$, $\{c_i,c_j\}=0=\{c_i^\dag,c_j^\dag\}$. $A,B,C$ are $N\times N$ matrices. We assume without loss of generality that $B$ and $C$ are antisymmetric, and Einstein summation over repeated indices is implied. Writing $f_i^\dag=c_{i}^\dag$ and $f_{i+N}^\dag=c_i$ for $i=1,\ldots,N$, the Hamiltonian can be rewritten as
\begin{align}\label{eq:generalquadratic}
	H=\frac{1}{2} F_{ij} f_i^\dag f_j+\frac{1}{2}\textrm{Tr}\,A,
\end{align}
where the $2N\times 2N$ matrix $F$ reads in block form
\begin{align}\label{eq:Fmat}
	F=\left(\begin{array}{cc}A& B\\ -C&-A^T\end{array}\right).
\end{align}
In case $H$ is Hermitian we have $A=A^\dag$ and $C=B^*=-B^\dag$. For the Ising model, the relevant Hamiltonian will be real symmetric.
\paragraph{Diagonalization of a quadratic Hamiltonian (real symmetric case).}
Consider a Hamiltonian \eqref{eq:generalquadratic} with $F$ given by \eqref{eq:Fmat} and $F$ real symmetric, which implies real symmetric $A$ and real antisymmetric $C=B$.
$F$ has particle-hole symmetry $\sigma^x F\sigma^z=-F$, which means its eigenvalues go in pairs $\pm \epsilon_k$ with $\epsilon_k\geq 0$. We assume for simplicity that no eigenvalue vanishes. The commutators
\begin{align}\label{eq:comm}
	[c_j\pm c_j^\dag,H]=(A_{jm}\mp B_{jm})(c_m\mp c_m^\dag)
\end{align}
suggest finding new fermionic operators with the Ansatz
\begin{align}
	b_k=v_{kj}^+ \frac{c_j+c_j^\dag}{2}+v_{kj}^- \frac{c_j-c_j^\dag}{2}.
\end{align}
Iterating \eqref{eq:comm} yields
\begin{align}
	[[c_j\pm c_j^\dag,H],H]=(A\mp B)_{jl}(A\pm B)_{lm}(c_m\pm c_m^\dag),
\end{align}
so $[b_k,H]=\epsilon_k b_k$ provided the $v_{kj}^+$ are obtained from the normalized eigenvectors of $(A-B)(A+B)$, with eigenvalues $\epsilon_k^2$, while the $v_{kj}^-$ are the matrix elements of $V_-=D^{-1/2}V_+(A-B)$, where $D$ is the diagonal matrix $D=\textrm{diag}(\epsilon_1^2,\ldots,\epsilon_N^2)$ and the $v_{kj}^+$ are the matrix elements of $V_+$. 

Solving the eigenvalue problem described above ensures that the $b_k,b_k^\dag$ satisfy the proper anticommutation relations $\{b_k,b_{k'}^\dag\}=\delta_{k,k'}$, $\{b_k,b_{k'}\}=0=\{b_k^\dag,b_{k'}^\dag\}$. The final diagonalized Hamiltonian reads
\begin{align}
	H=\sum_{k=1}^N \epsilon_k \left(b_k^\dag b_k-\frac{1}{2}\right)
\end{align}
with $\epsilon_k\geq 0$. In our case no $\epsilon_k$ vanishes in the ground state sector. Therefore the ground state $\ket{\psi_0}$ is the vacuum of the $b_k$, $b_k\ket{\psi_0}=0$, its energy is given by $E_0=-\frac{1}{2}\sum_k \epsilon_k$, and the gap $\Delta E$  is the smallest single particle energy $\epsilon_k$.
\paragraph{Jordan-Wigner mapping.}
The Jordan-Wigner mapping transforms spin-$1/2$ operators into fermionic ones, through
\begin{align}\label{eq:JordanWigner1}
	\sigma_j^x+\ci \sigma_j^y&= 2c_j^\dag\exp\left(\ci \pi \sum_{l=1}^{j-1}c_l^\dag c_l\right), \\\label{eq:JordanWigner2}
	\sigma_j^z&=2c_j^\dag c_j-1.
\end{align}
\paragraph{Product of exponentials.} With $F,F'$ of the form \eqref{eq:Fmat}:
 \begin{align}\label{eq:productofexp}
	\exp\left(\frac{1}{2}F_{ij}f_i^\dag f_j\right)\exp\left(\frac{1}{2}F'_{ij}f_i^\dag f_j\right)=\exp \left(\frac{1}{2}\log(e^F e^{F'})_{ij} f_i^\dag f_j\right).
\end{align}
Combined with the Jordan-Wigner mapping, this will allow to express the Ising transfer matrix \eqref{eq:generalIsingTM} as the exponential of a free fermion Hamiltonian \eqref{eq:generalquadratic}.
\subsection{Homogeneous Ising model}
For convenience, let us first perform a rotation in spin space which sends $\sigma^{x,z}\to \sigma^{z,x}$ and $\sigma^y \to -\sigma^y$. Denote by $U$ the corresponding unitary operator. Making use of the Jordan-Wigner transformation \eqref{eq:JordanWigner1},\eqref{eq:JordanWigner2} we obtain
\begin{align}
	U\mathcal{T} U^\dag= C \exp\left(K^* \sum_{j=1}^N 2c_j^\dag c_j-1\right) \exp\left(K' \sum_{j=1}^N(c_j^\dag-c_j)(c_{j+1}^\dag+c_{j+1})\right).
\end{align}
In the following and to lighten notations, we do not write down the corresponding unitaries $U,U^\dag$, but nevertheless keep the spin rotated basis. 
The previous transfer matrix has the disadvantage of not being symmetric. For this reason, it is customary to introduce its symmetrized counterpart
\begin{align}
	\mathcal{T}_{\rm s}=C \exp\left(\frac{K'}{2} \sum_{j=1}^N(c_j^\dag-c_j)(c_{j+1}^\dag+c_{j+1})\right)\exp\left(K^* \sum_{j=1}^N 2c_j^\dag c_j-1\right) \exp\left(\frac{K'}{2} \sum_{j=1}^N(c_j^\dag-c_j)(c_{j+1}^\dag+c_{j+1})\right)
\end{align}
Care must be taken with boundary conditions for both $\mathcal{T}$ and $\mathcal{T}_{\rm s}$. In this work we impose periodic boundary conditions for the spins, which implies $c_{N+1}^\dag=(-1)^{\hat{N}+1}c_1^\dag$, where $\hat{N}=\sum_{j=1}^N c_j^\dag c_j$ is the fermion number operator. Since both transfer matrices commute with $(-1)^{\hat{N}}$ this defines two sectors. Translation invariance means one can consider Fourier modes
\begin{align}
	c_k^\dag=\frac{1}{\sqrt{N}}\sum_{j=1}^N e^{\ci k j}c_j^\dag,
\end{align}
where the $e^{\ci k}$ are the $N$ possible roots of $(-1)^{\hat{N}+1}=\pm 1$, depending on sector. $k\in [-\pi,\pi]$ is momentum. Hence we obtain
\begin{align}
	\mathcal{T}=C\exp\left(K^* \sum_k (2 c_k^\dag c_k-1)\right)\exp\left(K'\sum_{k}2 \cos k c_k^\dag c_k-\ci \sin k[c_k^\dag c_{-k}^\dag+c_{k}c_{-k}]\right)
\end{align}
and a similar formula for $\mathcal{T}_{\rm s}$. Before making use of \eqref{eq:productofexp}, notice each term with momentum $k$  commutes with all $k'\neq k,-k$ in both sums. This means diagonalization essentially reduces to a $4\times 4$ matrix problem. The relevant matrices are
\begin{align}
	f(k)=2K^*\left(\begin{array}{cccc}1 & 0 &0 &0\\
	0&1&0&0 \\
	0&0&-1&0\\
	0&0&0&-1
	\end{array}\right)
	\qquad,\qquad 
	f'(k)=2K'\left(
	\begin{array}{cccc}
		\cos k & 0 & 0 &-\ci \sin k \\
		0& \cos k &\ci \sin k & 0\\
		0&-\ci \sin k & -\cos k & 0 \\
		\ci \sin k &0 &0 &-\cos k 
	\end{array}
	\right)
\end{align}
which reduces further to a $2\times 2$ problem, similar to the Bogoliubov-de Gennes formalism. The doubly degenerate eigenvalues $\pm \epsilon_k$ of $h(k)=\log [e^{f'(k)/2}e^{f(k)}e^{f'(k)/2}]$ satisfy the equation
\begin{align}\label{eq:coshepsilon}
\cosh \epsilon_k=\cosh 2K' \cosh 2K^*+\sinh 2K' \sinh 2K^* \cos k.
\end{align}
In the end the symmetric transfer matrix can be written as
\begin{align}
	\mathcal{T}_{\rm s}=C \exp\left(-\sum_k \epsilon_k \left[b_k^\dag b_k-\frac{1}{2}\right]\right),
\end{align}
where $b_k=\cos \theta_k c_k-\sin \theta_k c_{-k}^\dag$ for some angles which are obtained from the eigenvectors of $h(k)$, see \cite{SchultzMattisLieb}. By convention $\epsilon_k \geq 0$.

The ground state (eigenvector of $\mathcal{T}_{\rm s}$ corresponding to the largest eigenvalue) can be shown to lie in the sector $(-1)^{\hat{N}}=1$. Hence the momenta are quantized --recall $N$ is  even-- as $(2m+1)\pi/N$ for $m\in \{-N/2,-N/2+1,\ldots,N/2-1\}$. All results presented in the main text correspond to this ground state sector. 
\paragraph{Critical point and Fermi speed.} One can easily check from \eqref{eq:coshepsilon} that if $K'=K^*$, then the gap closes near $k=\pm\pi$. This is the well known criticality condition. Treating momentum as a continuous variable we find the linear behavior 
\begin{align}
	\epsilon_{k+\pi}\simeq \sinh (2K')\, k
\end{align} 
for small $k$
This allows to identify the Fermi speed that scales the low-energy conformal spectrum \cite{Yellowbook,Mussardo}, as $v_F=\sinh 2K'=\sinh 2K^*$.
\subsection{Inhomogeneous Ising model}
The treatment of the inhomogeneous model is very similar, the only significant difference being the loss of translation invariance, which makes the determination of the eigenmodes more complicated. The transfer matrix can be brought to the form
\begin{align}
	\mathcal{T}_{\rm s}=C \exp\left(\frac{1}{4}F'_{ij} f_i^\dag f_j\right)
	\exp\left(\frac{1}{2}F_{ij}f_i^\dag f_j\right)
	\exp\left(\frac{1}{4}F'_{ij}f_i^\dag f_j\right),
\end{align}
where $C=\prod_{j=1}^N \sqrt{2 \sinh 2K_j}$. 
The relevant $2N\times 2N$ matrices are
\begin{align}
	F&=2\,\textrm{diag}(K_1^*,\ldots,K_N^*,-K_1^*,\ldots,-K_N^*),\\
	F'&=\left(\begin{array}{cc}A'&B'\\-B'&-A'\end{array}\right),
\end{align}
where $A',B'$ have elements
\begin{align}
	A'_{ij}&=K'_{j+1/2}\delta_{i,j+1}+K'_{i+1/2}\delta_{i+1,j}-K'_{1/2}(\delta_{iN}\delta_{j1}+\delta_{jN}\delta_{i1}), \\
	B'_{ij}&=K'_{j+1/2}\delta_{i,j+1}-K'_{i+1/2} \delta_{i+1,j} - K'_{1/2}(\delta_{iN}\delta_{j1}-\delta_{jN}\delta_{i1}),
\end{align}
in the ground state sector. Finally using \eqref{eq:productofexp} we obtain
\begin{align}
	\mathcal{T}_{\rm s}=C \exp\left(\frac{1}{2} G_{ij}f_i^\dag f_j\right),
\end{align}
with $e^G=e^{F'/2}e^F e^{F'/2}$. It is worth noting that the matrices $F,F'$ have a simple quasi-tridiagonal structure, which allows to determine also $e^G$ in closed form. In practice, we simply evaluated $G$ numerically to get the results discussed in the main text. The necessary steps to perform the diagonalization are explained in appendix \ref{app:freefermions}.
\paragraph{Spin-spin correlations.}
In the infinite cylinder geometry $M\to\infty$ the classical spin-spin correlations are time-independent, and follow from the quantum ground state $\ket{\psi_0}$. Recalling that after spin rotation the two eigenvalues of the $\sigma_j^x$ operator encode the two classical spin states $\pm 1$ at site $j$, we obtain
\begin{align}
	\braket{\sigma(m,0)\sigma(m+\ell)}&=\braket{\psi_0|\sigma_m^x \sigma_{m+\ell}^x|\psi_0}\\
	&=\det_{m\leq i,j\leq m+\ell-1} \left(\braket{\psi_0|(c_i^\dag-c_i)(c_{j+1}^\dag+c_{j+1})|\psi_0}\right)
\end{align}
To get the result we used the Jordan-Wigner mapping, the identity $e^{\ci \pi c_j^\dag c_j}=(c_j^\dag+c_j)(c_j^\dag-c_j)$, and Wick's theorem. Noticing that the matrix $V_-^T V_+$ contains all necessary correlators, we arrive at the determinant formula
\begin{align}
	\braket{\sigma(m,0)\sigma(m+\ell)}=\det_{m\leq i,j\leq l+\ell-1}\left((V_-^T V_+)_{i,j+1}\right),
\end{align}
which can be evaluated numerically using standard linear algebra routines.
\pagebreak
\section{A Coulomb gas integral}
\label{app:cg}
In this appendix we compute the leading asymptotic behavior of the Coulomb gas integral 
\begin{align}\label{eq:cglesstypical}
D_N&=\frac{2^{N^2}}{N!}\left(\frac{N}{\alpha}\right)^N \int_{\mathbb{R}^N} dx_1,\ldots dx_N \prod_{1\leq j<k\leq N}\sinh^2(x_j-x_k) \prod_{j=1}^N e^{-NV(x_j)},
\end{align}
relevant to the discussion in section \ref{sec:partitionfunction}, see \eqref{eq:coulombgasintegral}. $V$ is given by
\begin{align}
V(x)=\frac{2 x}{\tau}\left[\kappa \,\Theta(x)+(\kappa-\pi)\, \Theta(-x)\right],
\end{align}
with the notations $\alpha=\tau/\delta$, $\lambda-\mu=\kappa/\delta$. For simplicity we assume $\tau>0$.
An alternative formulation can be obtained from the change of variables $u=\tanh x$:
\begin{align}\label{eq:cgtypical}
	D_N&=\frac{2^{N^2}}{N!}\left(\frac{N}{\alpha}\right)^N  \int_{[-1,1]^N} du_1\ldots du_N \prod_{1\leq j<k\leq N}(u_j-u_k)^2 \prod_{j=1}^N e^{-N W(u_j)}
\end{align} 
where
\begin{align}
	W(u)&=\log(1-u^2)+V(\textrm{arctanh}\, u)\\
	&=\log(1-u^2)+\frac{1}{\tau}\log \frac{1+u}{1-u}\left[\kappa\,  \Theta(u)+(\kappa-\pi)\, \Theta(-u)\right].
\end{align}
The multiple integral \eqref{eq:cgtypical} is not necessarily more convenient than \eqref{eq:cglesstypical}, however its asymptotic behavior happens to be well studied, see e.g. \cite{Forrester} for a general discussion of such Coulomb gas/Random matrix integrals. To the leading order $D_N\simeq e^{-N^2 (E[\rho]-\log 2)}$ where
\begin{align}
	E[\rho]=-\frac{1}{2}\int du dv \rho(u)\rho(v)\log(u-v)^2+\int du \rho(u)W(u)
\end{align}
is a continuum version of the energy appearing when rewriting the integrant as a Boltzmann weight. The density $\rho$ encodes the most likely positions for the integration variables in the $N\to\infty$ limit, it has nothing to do with the density discussed in section \ref{sec:6vdwfree}. For our problem it has support $[u_0,u_1]$, and solves the linear integral equation
\begin{align}\label{eq:lie}
	\int_{u_0}^{u_1} d v \rho(v)\log(u-v)^2=W(u)+C
\end{align}
for any $u\in [u_0,u_1]$. $C$ is a numerical constant. The closed form solution reads \cite{Forrester}
\begin{align}
	\rho(u)&=\frac{\sqrt{(u-u_0)(u_1-u)}}{2\pi^2}\int_{u_0}^{u_1} \frac{W'(u)-W'(w)}{u-w} \frac{dw}{\sqrt{(w-u_0)(u_1-w)}},
\end{align}
where $u_0,u_1$ are determined from the conditions
\begin{align}
	\int_{u_0}^{u_1} dw\frac{ W'(w)}{\sqrt{(w-u_0)(u_1-w)}}=0 \qquad,\qquad \int_{u_0}^{u_1}dw \frac{ w W'(w)}{\sqrt{(w-u_0)(u_1-w)}}=2\pi.
\end{align}
We find
\begin{align}
	u_0&=-\frac{\sin \tau\sin \kappa}{1+\cos \tau \cos \kappa}, \\
	u_1&=\frac{\sin \tau\sin \kappa}{1-\cos \tau \cos \kappa}.
\end{align}
Notice that $-1<u_0\leq u_1<1$ for $\tau<\max(\kappa,\pi-\kappa)$, which corresponds to the physical case of positive Boltzmann weights. After some computations we obtain
\begin{align}
	\rho(u)=\frac{\textrm{arccosh}\left[\cos \kappa \cos \tau \,\textrm{sgn}(u)+\frac{\sin \kappa \sin \tau}{|u|}\right]}{\pi \tau(1-u^2)}.
\end{align}
Using the above solution for the density $\rho$, the linear integral equation \eqref{eq:lie} and $W(0)=0$ we find
\begin{align}
	E[\rho]=\frac{1}{2}\int_{u_0}^{u_1} dx \rho(x)\left[W(x)-\log x^2\right].
\end{align}
After cumbersome computations, we finally obtain the leading order asymptotics
\begin{align}
	-\frac{\log D_N^\gamma}{N^2}&\to  E[\rho]-\log 2\\
	&=\frac{\zeta(3)}{2\tau^2}+\frac{\textrm{Re}\left\{Li_3(e^{2\ci(\kappa+\tau)})+Li_3(e^{2\ci(\kappa-\tau)})-2 Li_3(e^{2\ci \kappa})-2 Li_3(e^{2\ci \tau}) \right\}}{4\tau^2},
\end{align}
where $Li_s(z)=\sum_{k=1}^{\infty} \frac{z^k}{k^s}$. Recalling $\kappa=\delta(\lambda-\mu)$, $\tau=\delta \alpha$, this can be shown to equal
\begin{align}
	-\int_0^1 dx \int_0^x \log\left[ \sin \delta(\lambda(x)-\lambda(y))\sin \delta(\mu(x)-\mu(y))\right]
	+\int_0^1 dx \int_0^1 dy \log \sin \delta(\lambda(x)-\mu(y)),
\end{align}
consistent with the claims made in section \ref{sec:partitionfunction}, with in particular \eqref{eq:nice}  restated as
\begin{align}
	D_N^\gamma \simeq \delta^N \frac{\prod_{1\leq j<k\leq N} \sin \delta(\lambda_j-\lambda_k)\sin \delta(\mu_j-\mu_k)}{\prod_{j,k=1}^N \sin \delta(\lambda_j-\mu_k)}.
\end{align}
\end{appendix}
\pagebreak
\bibliography{biblio.bib}

\end{document}